\documentclass[%
 reprint,
nofootinbib,
 amsmath,amssymb,
 aps, superscriptaddress
]{revtex4-2}
\usepackage[T1]{fontenc} 

\usepackage{graphicx}
\usepackage{dcolumn}
\usepackage{bm}
\usepackage{graphicx}

\usepackage[bookmarks=true,colorlinks=true,linkcolor=magenta, urlcolor=blue, citecolor=blue]{hyperref}
\usepackage[normalem]{ulem}
\usepackage[section]{placeins}
\usepackage{amsmath}
\usepackage{booktabs}
\usepackage{multirow}
\usepackage{xcolor}
\usepackage[utf8]{inputenc}

\usepackage{amsmath}
\usepackage{amstext}
\usepackage{amssymb}
\usepackage{amsfonts}
\usepackage{amsthm}
\usepackage{mathtools} 

\usepackage{physics}
\usepackage{enumitem}
\usepackage{bbm}

\usepackage{caption}
\usepackage{subcaption}
\def\[#1\]{\begin{align}#1\end{align}}
\newcommand{\subalign}[1]{\begin{aligned}[t]#1\end{aligned}}

\newcommand{\C}{\mathcal{C}}

\renewcommand{\O}{\mathcal{O}}

\newcommand{\one}{\mathbbm{1}}
\renewcommand{\pmat}[1]{\begin{pmatrix}#1\end{pmatrix}}

\renewcommand{\ket}[1]{|#1\rangle}
\renewcommand{\braket}[2]{\langle#1|#2\rangle}
\renewcommand{\mel}[3]{\langle#1|#2|#3\rangle}

\RenewDocumentCommand{\v}{s m}{%
  \IfBooleanTF{#1}%
    {\bm{\hat{#2}}}%
    {\bm{#2}}%
}
\renewcommand{\dd}[2]{\frac{d #1}{d #2}}
\NewDocumentCommand{\D}{s m}{%
  \IfBooleanTF{#1}%
    {\mathrm{d}#2}%
    {\mathrm{d}#2\,}%
}

\newcommand{\braa}[1]{\langle\langle #1 |}
\newcommand{\kett}[1]{| #1 \rangle\rangle}
\newcommand{\braakett}[2]{\langle\langle #1 | #2 \rangle\rangle}

\newcommand{\ot}{\otimes}

\makeatletter
\newcommand{\phantomlabel}[2]{
    \protected@write\@auxout{}{
        \string\newlabel{#2}{
            {\@currentlabel#1}{\thepage}
            {\@currentlabel#1}{#2}{}
        }
    }
    \hypertarget{#2}{}
}
\makeatother

\begin{document}

\title{Continuous symmetry breaking in adaptive quantum dynamics}

\author{Jacob Hauser}
\affiliation{Department of Physics, University of California, Santa Barbara, CA 93106, USA}

\author{Yaodong Li}
\affiliation{Department of Physics, Stanford University, Stanford, CA 94305, USA}
\affiliation{Department of Physics, University of California, Santa Barbara, CA 93106, USA}

\author{Sagar Vijay}
\affiliation{Department of Physics, University of California, Santa Barbara, CA 93106, USA}

\author{Matthew P. A. Fisher}
\affiliation{Department of Physics, University of California, Santa Barbara, CA 93106, USA}

\begin{abstract}
Adaptive quantum circuits, in which unitary operations, measurements, and feedback are used to steer quantum many-body systems, provide an exciting opportunity to generate new dynamical steady states.  We introduce an adaptive quantum dynamics with continuous symmetry where unitary operations, measurements, and local unitary feedback are used to drive ordering. In this setting, we find a pure steady state hosting symmetry-breaking order, which is the ground state of a gapless, local Hamiltonian.  We explore the dynamical properties of the approach to this steady state. We find that this steady-state order is fragile to perturbations, even those that respect the continuous symmetry.
\end{abstract}

\maketitle

\section{Introduction}\label{sec:introduction}
Characterizing the equilibrium phases and phase transitions of quantum many-body systems has been a highly successful program for understanding the important features of these systems~\cite{wen2007qft}.
However, the ensembles of states available to quantum systems at thermal equilibrium are quite special.
Out of equilibrium, the possibilities are much broader and an organizing principle is lacking.

Random quantum circuits composed of local unitary operators have proved a valuable model for studying the dynamics of entanglement and conserved quantities far from equilibrium~\cite{Nahum_2017, Nahum_2018, Zhou_2020, von_Keyserlingk_2018, fisher2022random}.
More recently, monitored quantum circuits have emerged as a rich setting for exploring non-equilibrium physics. These quantum circuits may include projective measurements and classical processing in addition to local unitary operators. They comprise a broad family of models that can be as minimally constrained as random unitary evolution or as finely specified as particular quantum algorithms.

Random unitary circuits generally produce a featureless ensemble of states. In contrast, monitored quantum circuits can protect an area-law-entangled steady-state phase from this infinite-temperature ``thermalizing'' phase~\cite{Skinner_2019, Li_2018, Li_2019}, or give rise to volume-law-entangled steady states that do not arise in thermal equilibrium~\cite{li2023entanglement}. It has been found that area-law phases arising in this context can host steady states equivalent to ground states of gapped Hamiltonians, such as those with broken discrete symmetries~\cite{Sang_2021} or with discrete topological order~\cite{Lavasani_2021A,Lavasani_2021B,lavasani2022monitored}. 

However, these interesting steady-state features are manifest only in quantities that are nonlinear in the density matrix. Therefore, unlike ordinary expectation values that are linear in the density matrix, they cannot be observed without multiple copies of the same state, which necessarily requires postselection. This fact, in combination with the no-cloning theorem~\cite{Wootters1982,Dieks1982} and the generally uncontrollable random outcomes of projective measurements, makes it exponentially difficult to access these nonlinear features experimentally.

For special circuit architectures, aspects of this postselection problem can be avoided~\cite{Noel2022,Ippoliti_2021,Ippoliti_2022,Lu_2021,Li_2022,Tikhanovskaya_2023}. It is also possible to tackle this postselection with brute force for sufficiently small system sizes~\cite{Koh2022}, but this is not a scalable solution. Another path -- which we take up here -- is to use local unitary feedback conditioned on local measurement results to target a pure state with desired properties. The resulting dynamics of physical observables thus takes the form of a local quantum channel. More generally, feedback could be used to target mixed steady states of interest. In all of these cases, one can further relax the second locality constraint and utilize the power of classical communication to strengthen this approach~\cite{piroli2021LOCC}.

Recent literature has explored the possibility of using such local feedback to target states of interest, and has studied dynamical \textit{absorbing} phase transitions that arise when the strength of the feedback is varied~\cite{Lesanovsky_2019, Roy_2020, Iadecola_2022, Buchhold_2022, Chertkov_2022, Friedman_2022, McGinley_2022, Ravindranath_2022, ODea_2022, Sierant_2022, Piroli_2022}. In these works, the target states have been trivial product states or area-law-entangled states; they either do not break a symmetry or break a discrete symmetry. These states can be related to ground states of gapped Hamiltonians with finite-depth unitary circuits. Another body of work has focused on the preparation of gapped, topologically ordered states~\cite{raussendorf2003cluster, RBH2005, satzinger2021realizing, tantivasadakarn2021longrange, bravyi2022adaptive, lu2022shortcut, iqbal2023topological, fossfeig2023experimental} and mixed states \cite{lu2023mixed} with long-range quantum entanglement.

In this work, we explore the possibility of using feedback to target a steady state that is the ground state of a \emph{gapless} Hamiltonian, with continuous symmetry breaking and/or long-range entanglement. We illustrate that this is possible with a simple example in which a pure steady state with ferromagnetic order and logarithmic scaling of the entanglement entropy with subsystem size can be achieved in the presence of $U(1)$ symmetry.\footnote{Intriguingly, the manifold of steady states has a higher $SU(2)$ symmetry, as we discuss further in Sec.~\ref{sec:results:baseline_dynamics:steady_state}.} However, we find that the feedback in our model is insufficient to stabilize the ordered steady state in the presence of competing single-site Pauli channels, regardless of whether the Kraus operators for the channel respect the $U(1)$ symmetry.  In this case, we numerically observe the emergence of short-ranged ferromagnetic correlations, with a correlation length that diverges as the strength of the Pauli channel goes to zero.  Viewing the Pauli channel as ``unrecorded'' single-site projective Pauli measurements provides a natural way to unravel this mixed-state evolution into pure-state quantum trajectories, which we numerically find to exhibit area-law scaling of the entanglement entropy. Throughout this work we discuss both the quantum channel dynamics, as well as the evolution of pure-state trajectories which are an unravelling of the channel.

The rest of this paper is organized as follows. In Sec.~\ref{sec:model} we describe a baseline model of dynamics whose steady state is our pure state of interest (Sec.~\ref{sec:model:baseline_dynamics}), we describe a more general perturbed model of dynamics (Sec.~\ref{sec:model:perturbed_dynamics}), and we explore the symmetries present in these models (Sec.~\ref{sec:model:symmetries}). We present our results in Sec.~\ref{sec:results}, beginning with a discussion of relevant observables (Sec.~\ref{sec:results:observables}). Then we explore the steady state properties and dynamical features approaching steady state for each of three cases: the baseline dynamics (Sec.~\ref{sec:results:baseline_dynamics}), the perturbed dynamics with $U(1)$ symmetry (Sec.~\ref{sec:results:U1_perturbations}), and the perturbed dynamics without $U(1)$ symmetry (Sec.~\ref{sec:results:general_perturbations}). Finally, we close in Sec.~\ref{sec:discussion} by summarizing our results, discussing their limitations, and posing questions regarding the capabilities of monitored circuits with feedback and what we can learn from the quantum channel superoperator itself.

\section{Model}\label{sec:model}
\subsection{Baseline dynamics}\label{sec:model:baseline_dynamics}
We consider a circuit architecture composed of SWAP measurements and $\sigma^z$ unitary operators applied to a one-dimensional chain of $L$ qubits.

The SWAP operator has two eigenspaces: the three triplet states $\ket{S = 1,S^z = m}$ (for $m=-1,0,1$) are even under SWAP and the singlet state $\ket{S = 0, S^z=0}$ is odd.
Consequently, measuring SWAP on two sites is equivalent to measuring the total spin on these sites. The projectors corresponding to these eigenspaces are:
\[
\Pi^+ &= \dyad{1,1}{1,1}+\dyad{1,0}{1,0}+\dyad{1,-1}{1,-1}\\
\Pi^- &= \dyad{0,0}{0,0}
\] 
with $\Pi^\pm = (1 \pm \text{SWAP} )/2$.

Our model consists of two-site conditional processes wherein:
\begin{enumerate}
    \item The SWAP operator is measured across two sites.
    \item If the result is $-1$, a $\sigma^z$ gate is applied to the first site. This pumps the local singlet into a triplet state:
    \[
    \ket{0,0} \mapsto \ket{1,0}.
    \]
    Otherwise, no feedback is applied.
\end{enumerate}
One circuit time step in our model corresponds to $L$ such elementary steps. In each elementary step, one of these measurements with feedback is applied to a randomly selected bond. We assume open boundary conditions. This circuit architecture is illustrated in Fig.~\ref{fig:baseline_architecture}.

We consider both the quantum trajectory dynamics, where the SWAP measurement results are recorded, and the trajectory-averaged quantum channel dynamics. The corresponding quantum channel is obtained by averaging over the possible measurement outcomes, so that one elementary time step is given by
\[\label{eq:baseline_channel}
\C(\rho) = \subalign{
\frac{1}{L-1} \sum_{i=1}^{L-1} \big[&\Pi^+_{i,i+1}\rho \Pi^+_{i,i+1}\\
&+ \sigma^z_i\Pi^-_{i,i+1}\rho \Pi^-_{i,i+1}\sigma^z_i \big]
}
\]
and $\C^L(\rho)$ advances $\rho$ by one circuit time step. Conversely, the circuit quantum trajectories define a particular unravelling of the channel.

The quantum channel perspective is natural for considering the dynamics of observables which are linear in the density matrix of a state. Each quantum trajectory depends both on the sequence of randomly selected measurement locations (denoted $\v{x}$) and on the history of measurement results (denoted $\v{m}$). Thus, when we average the expectation value of an observable $\O$ over all possible trajectories $\ket{\psi_{\v{x},\v{m}}}$, we find that:
\[
\overline{\expval{\O}} &= \sum_{\v{x},\v{m}}p_{\v{x}}p_{\v{m}} \mel{\psi_{\v{x},\v{m}}}{\O}{\psi_{\v{x},\v{m}}}\\
&= \sum_{\v{x},\v{m}}p_{\v{x}}p_{\v{m}}  \Tr[\O\dyad{\psi_{\v{x},\v{m}}}{\psi_{\v{x},\v{m}}}]\\
&= \Tr[\O\rho],
\]
where
\[
\rho = \sum_{\v{x},\v{m}}p_{\v{x}}p_{\v{m}}\dyad{\psi_{\v{x},\v{m}}}{\psi_{\v{x},\v{m}}}
\]
is a mixed state corresponding to the ensemble of trajectories, and its dynamics is given by Eq.~\ref{eq:baseline_channel}.
Here $p_{\v{m}}$ is determined by the Born rule, and $p_{\v{x}} = (L-1)^{-tL}$ for a circuit with $t$ time steps since each elementary step selects randomly among $L-1$ bonds.\footnote{In the perturbed model introduced in Sec.~\ref{sec:model:perturbed_dynamics}, $p_{\v{x}}$ also depends on which perturbing measurements occur in the circuit $\v{x}$ and the various probabilities of these perturbations.}

Since this quantum channel is linear, we can also view it as a matrix acting on a doubled Hilbert space (see Appendix \ref{appendix:doubled_hilbert_space} for more information) where
\[
\rho &\to \kett{\rho}\\
K\rho K^\dag &\to (K\ot K^*) \kett{\rho}.
\]
Performing this transformation, we arrive at
\[\label{eq:baseline_channel_matrix}
\C = \frac{1}{L-1}\sum_{i=1}^{L-1}\Big[(\Pi^{+}_{i,i+1})^{\ot 2}+(\sigma^z_i\Pi^{-}_{i,i+1})^{\ot 2}\Big].
\]
This matrix determines two important properties of the dynamics: the steady state order and the saturation time.

\begin{figure}[h!]
    \centering
    \includegraphics[width=\linewidth]{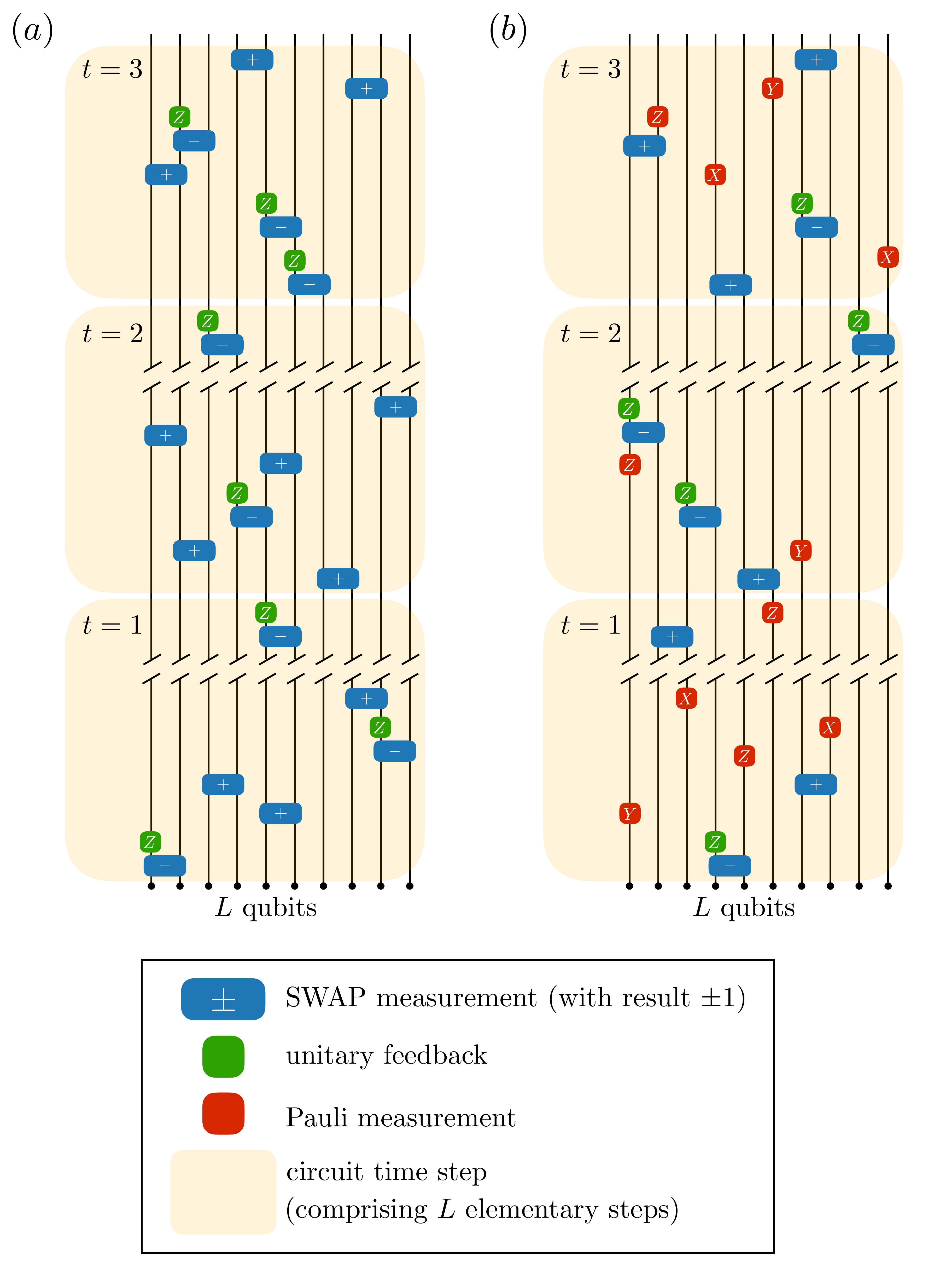}
    \caption{The baseline and perturbed circuit models we consider. In both cases, orange blocks indicate circuit time steps containing $L$ elementary steps. (a)~The baseline model consists only of SWAP measurements and, following $-1$ parity results, $\sigma^z$ unitary feedback. The SWAP measurements occur at random bonds with open boundary conditions. (b)~The perturbed model adds in the possibility of Pauli measurements. These are applied at random sites with probabilities $p_x$, $p_y$, and $p_z$ respectively (and SWAP measurements occur at random bonds with probability $p_s = 1- (p_x+p_y+p_z)$).}
    \phantomlabel{a}{fig:baseline_architecture}
    \phantomlabel{b}{fig:perturbed_architecture}
\end{figure}

\subsection{Perturbed dynamics}\label{sec:model:perturbed_dynamics}
We also study the impact of perturbing these dynamics with single-site Pauli measurements. Now, in each elementary time step, one of the following occurs:
\begin{enumerate}
    \item A SWAP measurement with $\sigma^z$ feedback is applied to a random bond with probability $p_s$.
    \item A single-site Pauli measurement occurs at a random site wherein
    \begin{enumerate}
        \item $\sigma^x$ is measured with probability $p_x$,
        \item $\sigma^y$ is measured with probability $p_y$, and
        \item $\sigma^z$ is measured with probability $p_z$,
    \end{enumerate}
    with $p_s+p_x+p_y+p_z = 1$.
\end{enumerate}
As in the unperturbed dynamics, a circuit time step comprises $L$ of these elementary steps. This circuit architecture is illustrated in Fig.~\ref{fig:perturbed_architecture}.

The corresponding perturbed quantum channel matrix is given by
\[
\C' = \subalign{\frac{1}{L-1}&\sum_{i=1}^{L-1}p_s\left[(\Pi^{+}_{i,i+1})^{\ot 2}+(\sigma^z_i\Pi^{-}_{i,i+1})^{\ot 2} \right]\\
+\frac{1}{2L}&\sum_{i=1}^L\left[(1-p_s)\one^{\ot2}+\sum_{\mu=x,y,z} p_\mu(\sigma_i^\mu)^{\ot2}\right].
}
\]

\subsection{Symmetries of our models}\label{sec:model:symmetries}
Since both SWAP and $\sigma^z$ commute with $Q = \sum_i S^z_i$, the baseline model has a strong $U(1)$ symmetry where $S^z_{\rm tot}$ is the corresponding conserved charge. We use the notion of ``strong symmetry'' in the sense that all Kraus operators of the channel commute with the symmetry generator, following Refs.~\cite{buvca2012note, lieu2020bosonic}. Thus, this symmetry remains when $\sigma^z$ measurements are introduced. As a result, the perturbed dynamics continues to have this $U(1)$ symmetry when $p_x=p_y=0$. However, it is broken when $p_x > 0$ or $p_y > 0$ since $\sigma^x$ and $\sigma^y$ do not commute with~$Q$.

\section{Results}\label{sec:results}
\subsection{Quantities of interest}\label{sec:results:observables}
Given our model's $U(1)$ symmetry when ${p_x = p_y = 0}$, quantities that diagnose spontaneous breaking of this symmetry are of particular interest. For this purpose, it is useful to consider the spin-spin correlation function $\mel{\psi}{\v{S}_i\cdot\v{S}_j}{\psi}$. Then we can diagnose symmetry breaking using the \emph{ferromagnetic susceptibility}:
\[\label{eq:susceptibility}
\chi = \frac{1}{L}\sum_{ij}\mel{\psi}{\v{S}_i\cdot\v{S}_j}{\psi}.
\]
We can also look for signatures of long-range order using the quantity $\mel{\psi}{\v{S}_1\cdot\v{S}_L}{\psi}$. In both cases, we are interested in the behaviour as $L \to \infty$. Since both $\chi$ and $\expval{\v{S}_1\cdot\v{S}_L}$ are linear in the density matrix, these quantities are accessible both in the quantum trajectory dynamics and in the quantum channel dynamics.

Additionally, we are interested in the entanglement dynamics of the baseline and perturbed models. Since entanglement entropies are nonlinear in the density matrix, their dynamics are visible only at the quantum trajectory level. Here, we focus on the half-chain von Neumann entropy:
\[
S(\rho_A) = -\Tr[\rho_A \log \rho_A]
\]
where $\rho_A = \Tr_B[\rho]$ with $A$ and $B$ indicating the left and right halves of the system, respectively.

When simulating quantum trajectories, it is necessary to pick an initial state. Unless otherwise stated, we use a half-filled N\'eel state:
\[
\ket{\psi_0} = \ket{{\uparrow}{\downarrow}{\uparrow}{\downarrow} \cdots {\uparrow}{\downarrow}}.
\]

\subsection{Baseline dynamics}\label{sec:results:baseline_dynamics}
\subsubsection{Steady state properties} \label{sec:results:baseline_dynamics:steady_state}
To study $U(1)$ symmetry breaking and entanglement dynamics in the baseline model, we first need to understand what steady states are present in this model.

We recall that the baseline dynamics is composed entirely of SWAP measurements with feedback. Consequently, a state $\ket{\psi}$ can be a steady state of the dynamics only if $\text{SWAP}_{i,i+1}\ket{\psi} = \ket{\psi}$ for all $i < L$. In this case, the state is unaffected by the SWAP measurement and no feedback is applied to change the state. We can actually make a stronger statement about $\ket{\psi}$: since the set of adjacent transpositions (operators like $\text{SWAP}_{i,i+1}$) generate the entire space of permutations, it follows that $\text{SWAP}_{ij}\ket{\psi} = \ket{\psi}$ for any $i$ and $j$.

Expressing the SWAP measurement in terms of total spin provides helpful physical intuition. We can write the SWAP operator on sites $i$ and $j$ as $\text{SWAP}_{ij} = \tfrac{1}{2}(\one_i \one_j+\v{\sigma}_i\cdot\v{\sigma}_j)$ where
\[
\v{\sigma}_i\cdot\v{\sigma}_j = \sigma^x_i\sigma^x_j+\sigma^y_i\sigma^y_j+\sigma^z_i\sigma^z_j = 4(\v{S}_i\cdot\v{S}_j),
\]
with $\hbar = 1$. Therefore, the state with even parity along each bond (and, consequently, between any two sites) will also maximize total spin. It follows that pure steady states along quantum trajectories are those states with maximal total spin. There are $L+1$ such states, labeled $\ket{\tfrac{L}{2},m}$ with $\abs{m} \leq \frac{L}{2}$. Because the space of states with maximal total spin is invariant under $SU(2)$ rotations, the emergent symmetry of the manifold of steady states is higher than the microscopic dynamical $U(1)$ symmetry of the model. Since there is one such steady state in each charge sector, the $U(1)$ symmetry dictates that there is a unique steady state if the initial state has fixed charge. 

One could worry that the quantum channel dynamics has additional mixed steady states that prevent the states $\ket{\tfrac{L}{2},m}$ from being attractive fixed points of the dynamics. However, we anticipate that these pure states are the unique steady states of both the trajectory and channel dynamics, and we find that this is consistent with our numerics. 

We can immediately calculate the ferromagnetic susceptibility and long-range order for these steady states. Rewriting the susceptibility in Eq.~\ref{eq:susceptibility} in terms of SWAP operators, we find that
\[
\chi &= \frac{1}{L}\sum_{i\neq j}\mel{\tfrac{L}{2},m}{\v{S}_i\cdot\v{S}_j}{\tfrac{L}{2},m} + \frac{3}{4}\\
&= \frac{1}{4L}\sum_{i\neq j}\mel{\tfrac{L}{2},m}{(2\text{SWAP}_{ij}-1)}{\tfrac{L}{2},m} + \frac{3}{4}
\]
in the steady states. Since any $\ket{\frac{L}{2},m}$ is totally symmetric, $\mel{\frac{L}{2},m}{\text{SWAP}_{ij}}{\frac{L}{2},m} = 1$ for all $i,j$ and
\[
\chi &= \frac{L(L-1)}{4L}+\frac{3}{4} = \frac{L+2}{4}.
\]
We conclude that the ferromagnetic susceptibility diverges in the thermodynamic limit for these steady states. Similarly, we find that $\expval{\v{S}_1\cdot\v{S}_L} = \frac{1}{4}$ for these steady states.

In Appendix \ref{appendix:entanglement} we show that, given an initial state with fixed charge $Q$, the steady-state bipartite entanglement entropy grows as
\[
S (\rho_A) &=  \subalign{&\frac{1}{2}\log L + \frac{1}{2}(\log(2\pi abn(1-n))+1) \\
&+ O\left(\frac{\log L}{L}\right)}\label{eq:entanglement_growth}
\]
in the baseline model as $L \to \infty$ with $q = \frac{Q}{L}$ held constant, and with $q$ satisfying $-\frac{1}{2} < q < \frac{1}{2}$. In Eq.~\ref{eq:entanglement_growth}, $a = \frac{|A|}{L}$ and $b=1-a$ are the fraction of spins in the two bipartitions respectively, and $n=q+\tfrac{1}{2}$ is the number of spins pointed up (in the $z$ direction).\footnote{The leading order contribution agrees with results for the ideal Bose gas~\cite{metlitski2011entanglement}. This is a sensible comparison since the spins in our one-dimensional model, although hard-core bosons, have a steady state which resembles that of a 1d ferromagnet, which has long-ranged order like the ground state of an ideal Bose gas.} This result agrees with our MPS simulations, as shown in Fig.~\ref{fig:baseline_entanglement_growth}.

\subsubsection{Approach to steady state} \label{sec:results:baseline_dynamics:approach}
Observables that are linear in the density matrix -- like the ferromagnetic susceptibility and long-range order -- are accessible in the quantum channel dynamics. Consequently, no such quantity can saturate more slowly than the rate at which the quantum channel reaches its steady state. In this way, the lifetime of the channel's longest-lasting non-steady mode sets a fundamental time scale for linear observables in both the quantum trajectory and quantum channel dynamics.

\begin{figure}[h!]
    \centering
    \begin{subfigure}[b]{\linewidth}
        \centering
        \includegraphics[width=\linewidth]{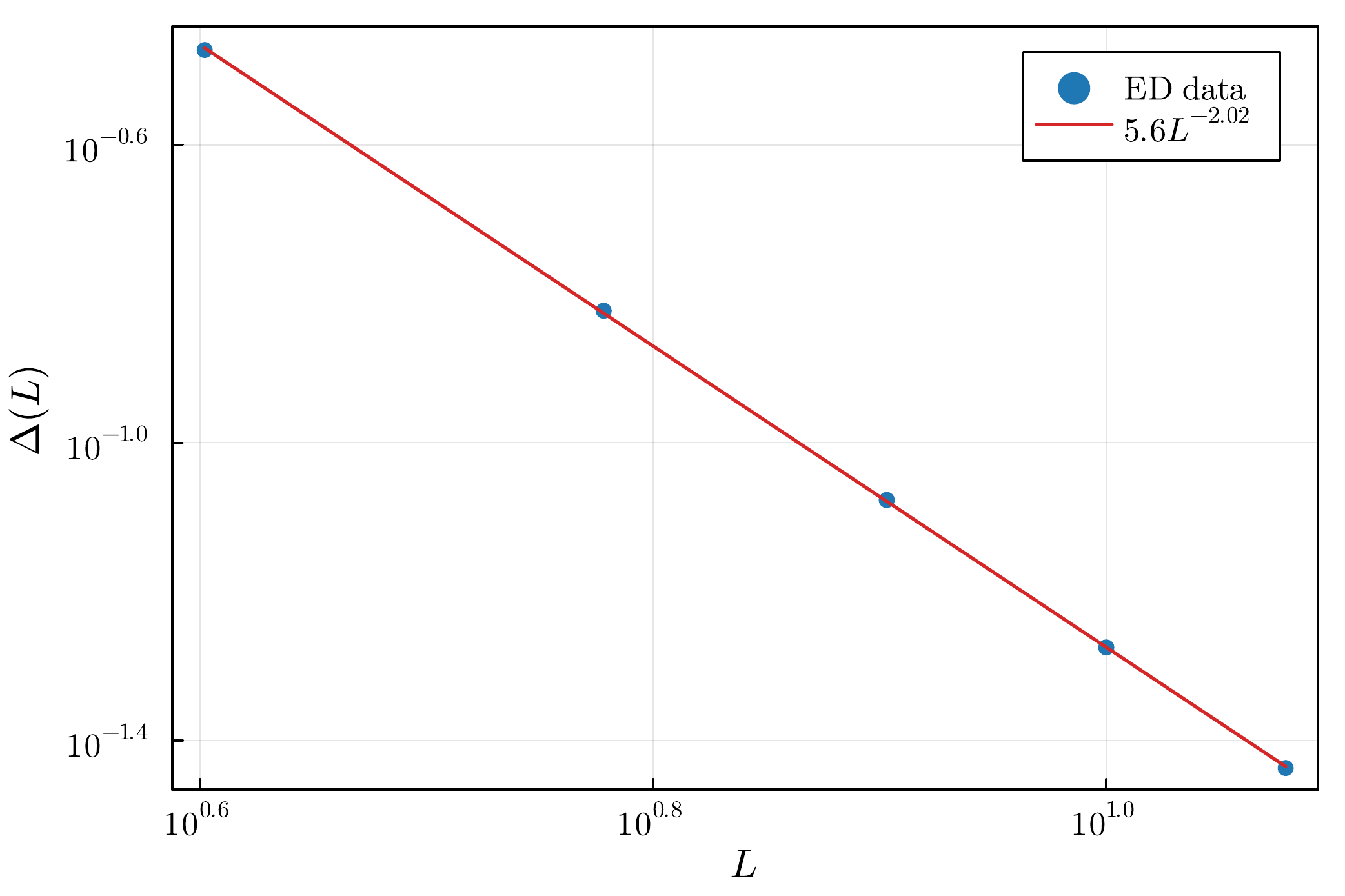}
        \caption{}
        \label{fig:baseline_channel_gap}
    \end{subfigure}
    \hfill
    \vspace{0.0em}
    \begin{subfigure}[b]{\linewidth}
        \centering
        \includegraphics[width=\linewidth]{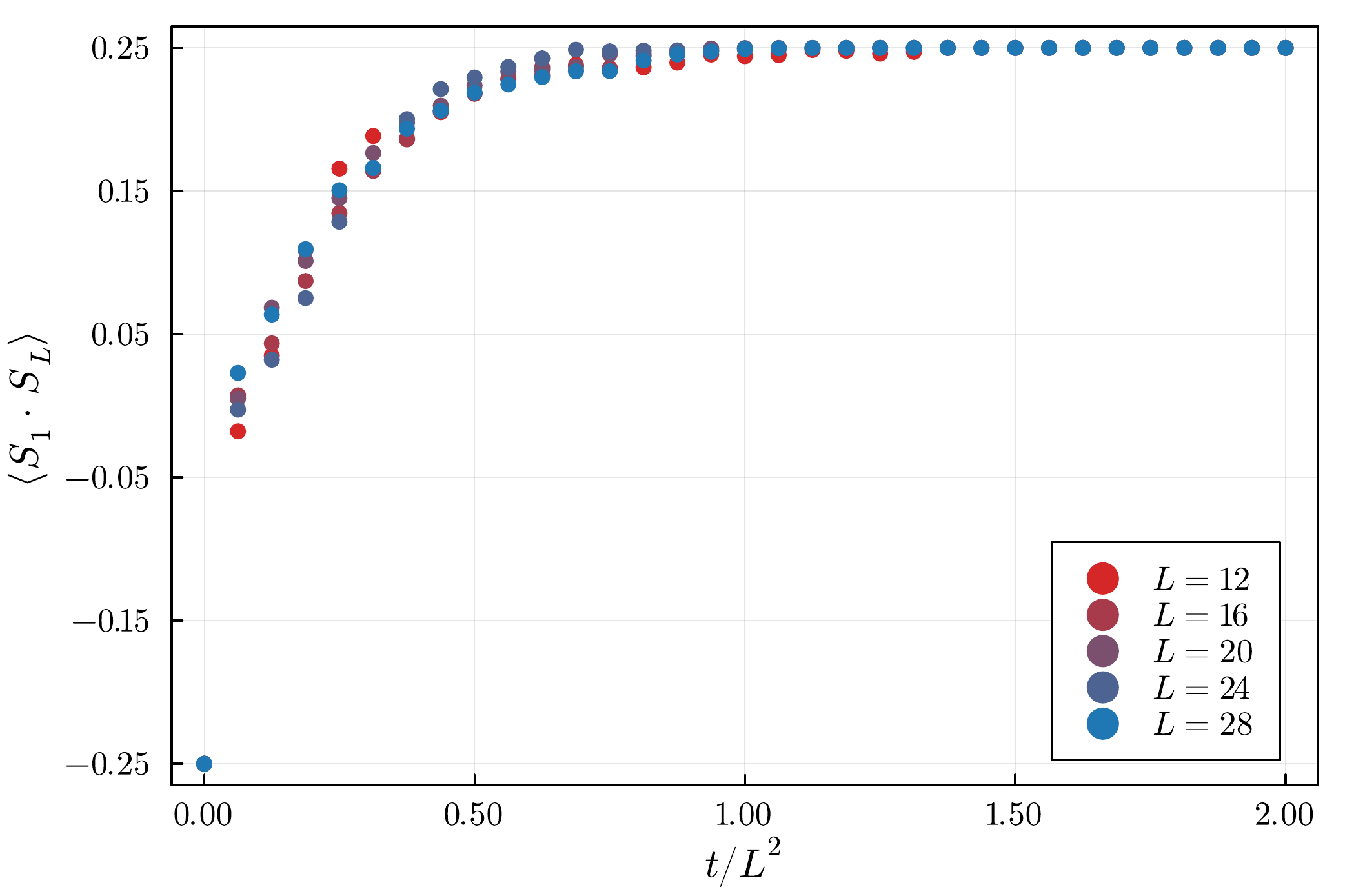}
        \caption{}
        \label{fig:baseline_SS_growth}
    \end{subfigure}
    \caption{(a)~Channel gap in a single charge sector over a range of system sizes for the baseline model (see Sec.~\ref{sec:results:baseline_dynamics}). We find that the gap shrinks with $L^{-z}$ where $z =2.02(1)$. (b)~Growth of $\expval{\v{S}_1\cdot\v{S}_L}$ over time for a range of system sizes in the baseline model. For each system size, the data are the average of $100$ quantum trajectories simulated using MPS tensor networks. We find that $\expval{\v{S}_1\cdot\v{S}_L}$ saturates in $O(L^2)$ time steps in the quantum trajectory approach, which is consistent with the gap scaling seen in the quantum channel approach.}
\end{figure}

As described in Appendix \ref{appendix:channel_gaps}, the lifetime of the longest-lasting non-steady mode of a channel $\mathcal{E}$ is set by the gap $\Delta$ in the corresponding non-Hermitian ``Hamiltonian'' $H = \one-\mathcal{E}$.\footnote{There may be exceptional cases where this does not apply. Because the channel matrix is not generally Hermitian, its eigenvectors are generally not mutually orthogonal. This allows initial states to have extensively large weight on non-steady modes that therefore take longer to decay. This is discussed in the context of Lindblad evolution in~\cite{Mori_2020}, for example. Having said all this, our MPS simulations of quantum trajectories suggest that the channel gap is representative of the time scales in our models.} In particular, the channel's saturation time is proportional to $\Delta^{-1}$. For the case of our baseline model, $\mathcal{E} = \C^L$. In Fig.~\ref{fig:baseline_channel_gap}, we plot the scaling of $\Delta$ with system size, calculated using exact diagonalization on the baseline channel matrix. We find results consistent with $\Delta(L) \propto L^{-2}$, meaning that $O(L^2)$ circuit time steps are required for linear observables to saturate. This is consistent with the intuition that conserved charges spread diffusively under random dynamics.

We can check this diffusive behaviour for larger system sizes by studying the saturation of $\expval{\v{S}_1\cdot\v{S}_L}$ in quantum trajectories using MPS simulations. These data are plotted for a range of system sizes in Fig.~\ref{fig:baseline_SS_growth}, and are consistent with the scaling suggested by the channel gap.

\begin{figure}[h!]
    \centering
    \includegraphics[width=\linewidth]{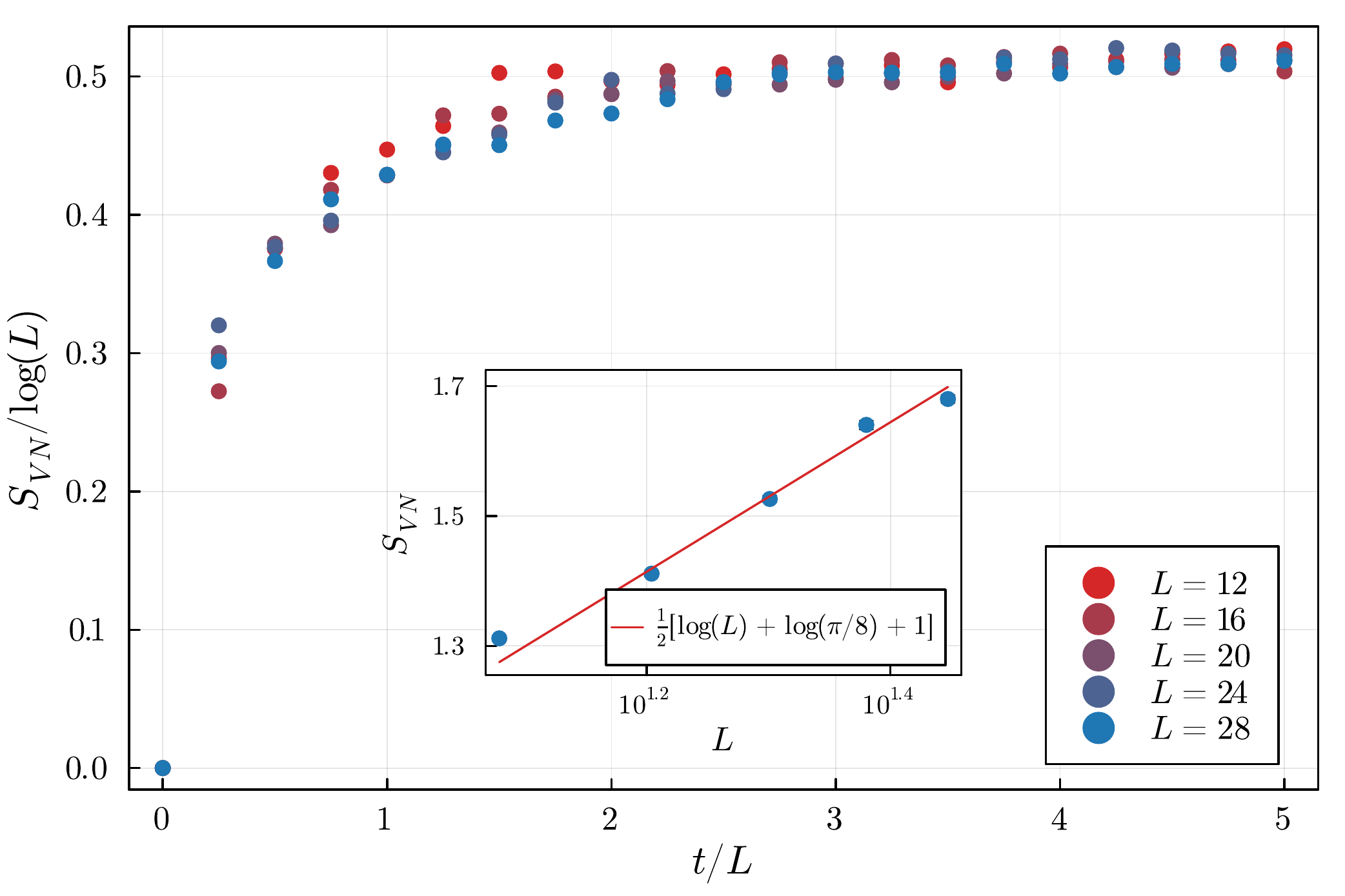}
    \caption{Half-chain entanglement growth with time over a range of system sizes for the baseline model (see Sec.~\ref{sec:results:baseline_dynamics}) starting from an initially unentangled N\'eel state. The data collapse that arises suggests that entanglement saturates in $O(L)$ time to a steady-state value that is logarithmic in system size. This logarithmic growth is visualized in the inset plot, where the steady-state entanglement is compared against the leading order analytical prediction given in Eq.~\ref{eq:entanglement_growth} (with $a=b=n=\frac{1}{2}$). For each system size in the main plot, the data are the average of $100$ quantum trajectories. Each point in the inset is the average of $1200$ samples, composed of $12$ measurements between $t=5L$ and $t=8L$ for each of the $100$ trajectories.}
    \label{fig:baseline_entanglement_growth}
\end{figure}

Of course, the channel gap has no implications for the dynamics of quantities like entanglement entropy that are nonlinear in the density matrix. Nevertheless, we can probe entanglement dynamics through MPS simulations of quantum trajectories. As illustrated in Fig.~\ref{fig:baseline_entanglement_growth}, we find that entanglement entropy saturates in $O(L)$ time in our baseline model.

\subsection{Perturbed dynamics with $U(1)$ symmetry} \label{sec:results:U1_perturbations}
As discussed in Sec.~\ref{sec:model:symmetries}, the perturbed model retains $U(1)$ symmetry when $p_x = p_y = 0$. We address this case in this Sec.~\ref{sec:results:U1_perturbations}, before taking up the general perturbed model in Sec.~\ref{sec:results:general_perturbations}.

\subsubsection{Steady state properties}\label{sec:results:U1_perturbations:steady_state}
When any single-site Pauli measurements are present in the perturbed model, the pure states $\ket{\frac{L}{2},m}$ are generally no longer steady states. In particular, when $p_x = p_y = 0$, these states are only steady states for $m = \pm\frac{L}{2}$. In other words, only $\ket{{\uparrow}{\uparrow}\cdots{\uparrow}}$ and $\ket{{\downarrow}{\downarrow}\cdots {\downarrow}}$ are steady states along trajectories. Since $Q$ is conserved in this case, quantum trajectories beginning with an initial state in a non-extremal charge sector will have no pure steady states.

\begin{figure}[h!]
    \centering
    \includegraphics[width=\linewidth]{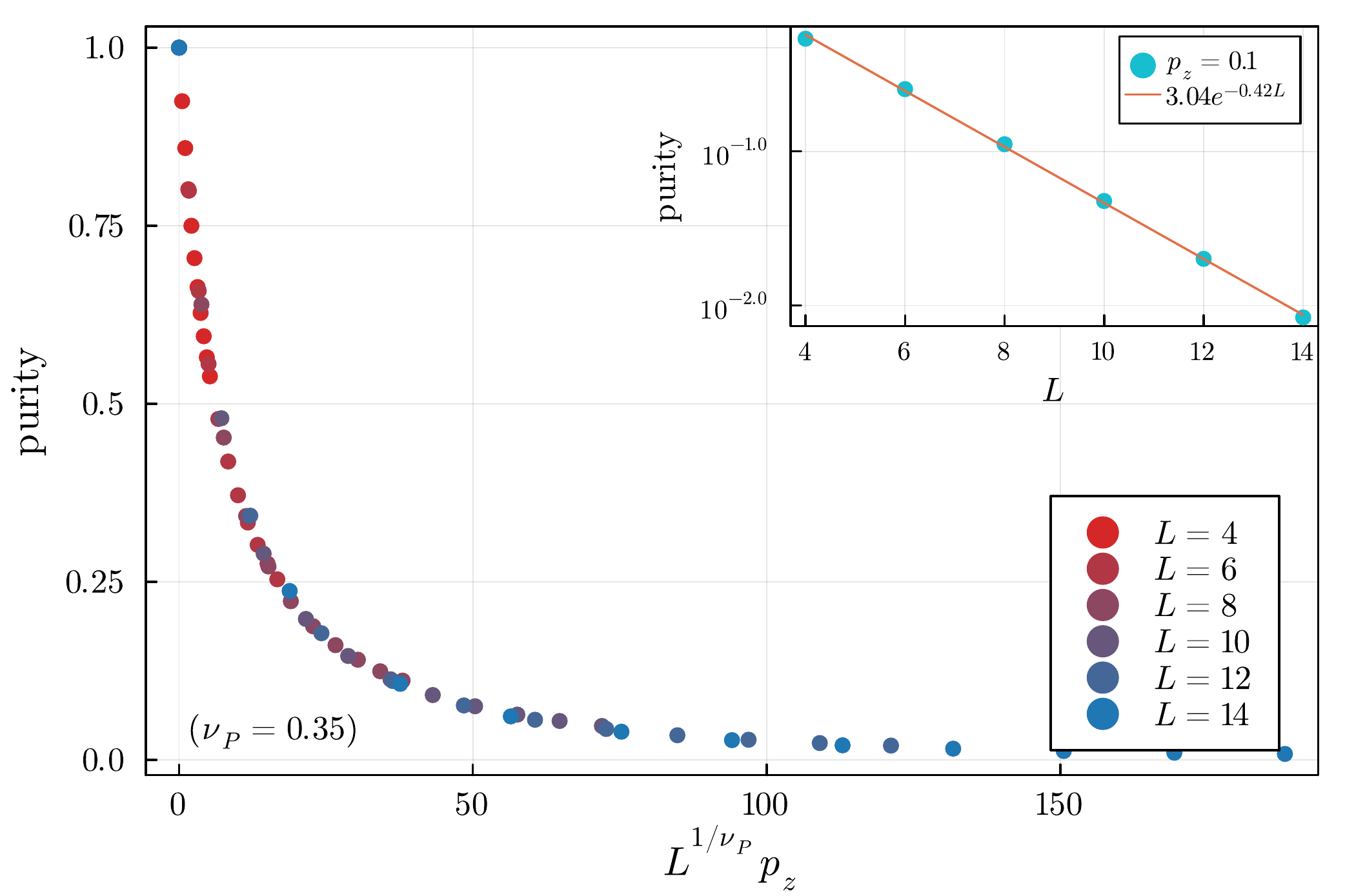}
    \caption{Steady state purity $P = \Tr[\rho^2]$ of the quantum channel (at half-filling) as a function of $L^{1/{\nu_P}} p_z$ in the perturbed model with $U(1)$ symmetry (see Sec.~\ref{sec:results:U1_perturbations}). We find that the data collapse as $P(L,p_z) = F_P(L^{1/\nu_P}p_z)$ when $\nu_P \approx 0.35$. When $p_z > 0$, the purity decays exponentially with system size, as illustrated for $p_z = 0.1$ in the inset.}
    \label{fig:U1_purity_scaling}
\end{figure}

Correspondingly, the quantum channel steady state is mixed in non-extremal charge sectors when both $p_s$ and $p_z$ are nonzero. In Fig.~\ref{fig:U1_purity_scaling}, we illustrate the decay of the purity
\[
P = \Tr(\rho^2)
\]
as $p_z$ is increased. In particular, when we plot $P(L, p_z)$ against the scaling variable $L^{1/{\nu_P}}p_z$ we find that the data collapse to the scaling form 
\[
P(L,p_z) = F_P(L^{1/{\nu_P}}p_z)
\]
when $\nu_P \approx 0.35$. At this stage, we do not have an understanding of this scaling.

Using exact diagonalization on the quantum channel matrix we find that these mixed steady states are not ordered. Using MPS simulations of quantum trajectories, we confirm this and also find that the trajectories have area-law entanglement at late times. In contrast, the mixture of all steady-state trajectories accessed in the quantum channel approach has a volume-law entropy, corresponding to the exponentially small purity illustrated in the inset of Fig.~\ref{fig:U1_purity_scaling}.

Although the dynamics has an ordered steady state only when no perturbation is actually applied (that is, when $p_z=0$), we can look for a scaling form like that of the purity. For this purpose, it is desirable to consider the scaling of $\expval{S_1^xS_L^x + S_1^yS_L^y}$ instead of $\expval{\v{S}_1\cdot\v{S}_L}$ because the model's $U(1)$~symmetry protects $\expval{S_1^zS_L^z}$ correlations that are unrelated to breaking of the symmetry and are independent of $p_z$. In particular, we show in Appendix~\ref{appendix:ZZ} that, for any $i \neq j$,
\[
\Tr[S_i^zS_j^z\rho(L,p_z)] = \frac{4Q^2-L}{4L(L-1)}\label{eq:ZZ_scaling}
\]
where $\rho(L,p_z)$ denotes the mixed steady state in a given charge sector. As ${L \to \infty}$, we can organize this as $\expval{S_i^zS_j^z} = \langle S_i^z\rangle\langle S_j^z \rangle + O(L^{-1})$.

\begin{figure}[h!]
    \centering
    \includegraphics[width=\linewidth]{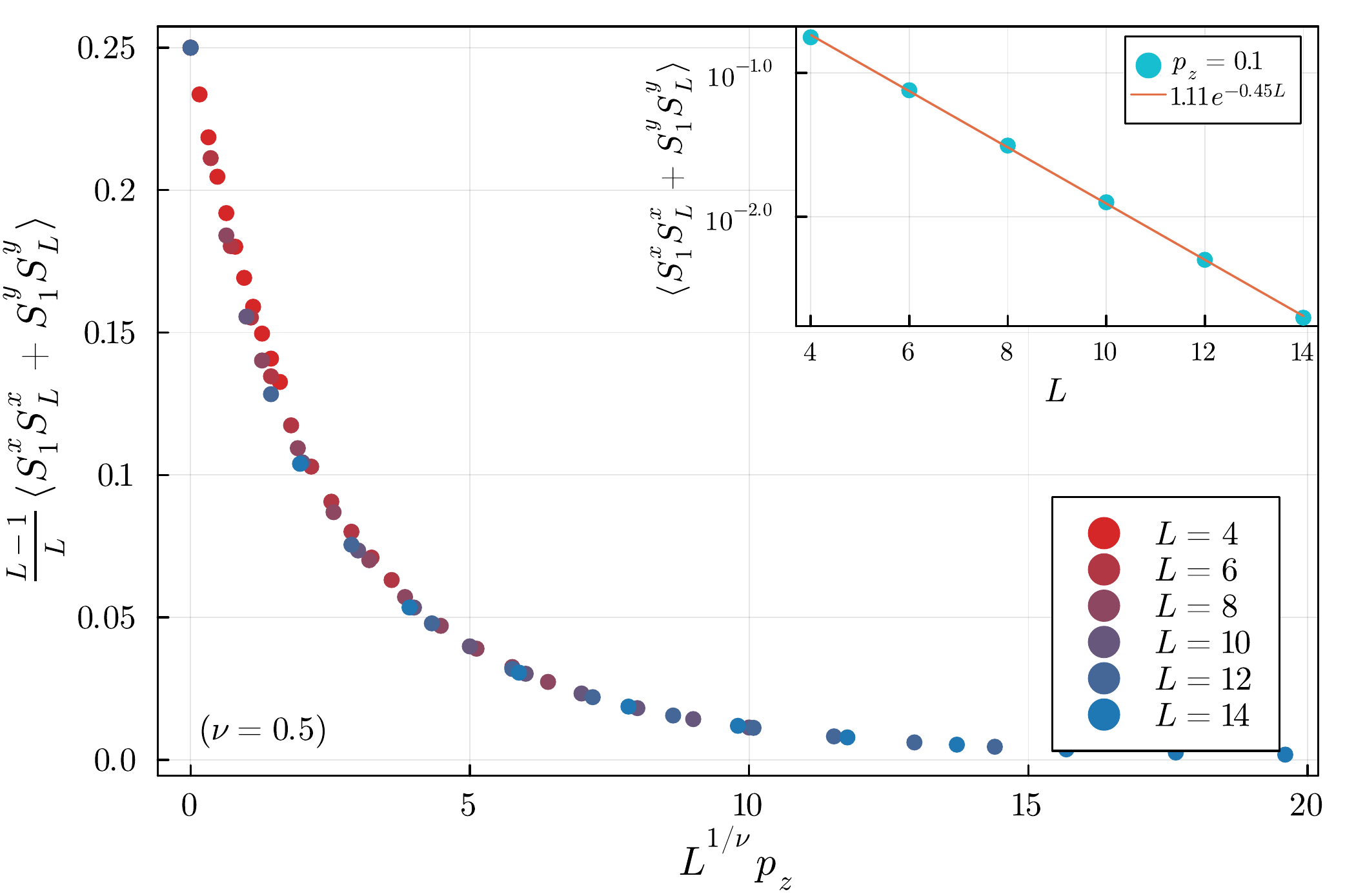}
    \caption{Steady state $\expval{S_1^xS_L^x+S_1^yS_L^y}$ of the quantum channel (at half-filling) as a function of $L^{1/\nu}p_z$ in the perturbed model with $U(1)$ symmetry (see Sec.~\ref{sec:results:U1_perturbations}). We find that the data collapse as $\frac{L-1}{L}\Tr[(S_1^xS_L^x+S_1^yS_L^y)\rho(L,p_z)] = F_{SS}(L^{1/\nu}p_z)$ when $\nu \approx 0.5$. The factor of $\frac{L-1}{L}$ corrects for the system-size dependence of $\expval{S_1^zS_L^z}$ given in Eq.~\ref{eq:ZZ_scaling} when $Q = 0$, and can be ignored in the thermodynamic limit. When $p_z > 0$, the value of $\expval{S_1^xS_L^x+S_1^yS_L^y}$ decays exponentially with system size as illustrated for $p_z = 0.1$ in the inset. $\expval{S_1^xS_L^x+S_1^yS_L^y}$ is a more suitable correlation function than $\expval{\v{S}_1\cdot\v{S}_L}$ for identifying a scaling form and diagnosing exponential decay because the latter has perturbation-independent $\expval{S_1^zS_L^z}$ correlations protected by the $U(1)$ symmetry.}
    \label{fig:U1_SS_scaling}
\end{figure}

Considering the $S_1^xS_L^x + S_1^yS_L^y$ correlations, we find that the data collapse to the scaling form
\[
\Tr[(S_1^xS_L^x + S_1^yS_L^y)\rho(L,p_z)] = F_{SS}(L^{1/\nu}p_z)
\]
with ${\nu \approx 0.5}$. This collapse is illustrated in Fig.~\ref{fig:U1_SS_scaling}. To make the collapse manifest at small system sizes, we scale $(S_1^xS_L^x + S_1^yS_L^y)$ by a factor of $\frac{L-1}{L}$, which approaches unity in the thermodynamic limit. In the inset of this figure we give an example of the exponential decay of $\expval{S_1^xS_L^x + S_1^yS_L^y}$ that arises when $p_z$ is nonzero.

\subsubsection{Approach to steady state} \label{sec:results:U1_perturbations:approach}
\begin{figure}[h!]
    \centering
    \includegraphics[width=\linewidth]{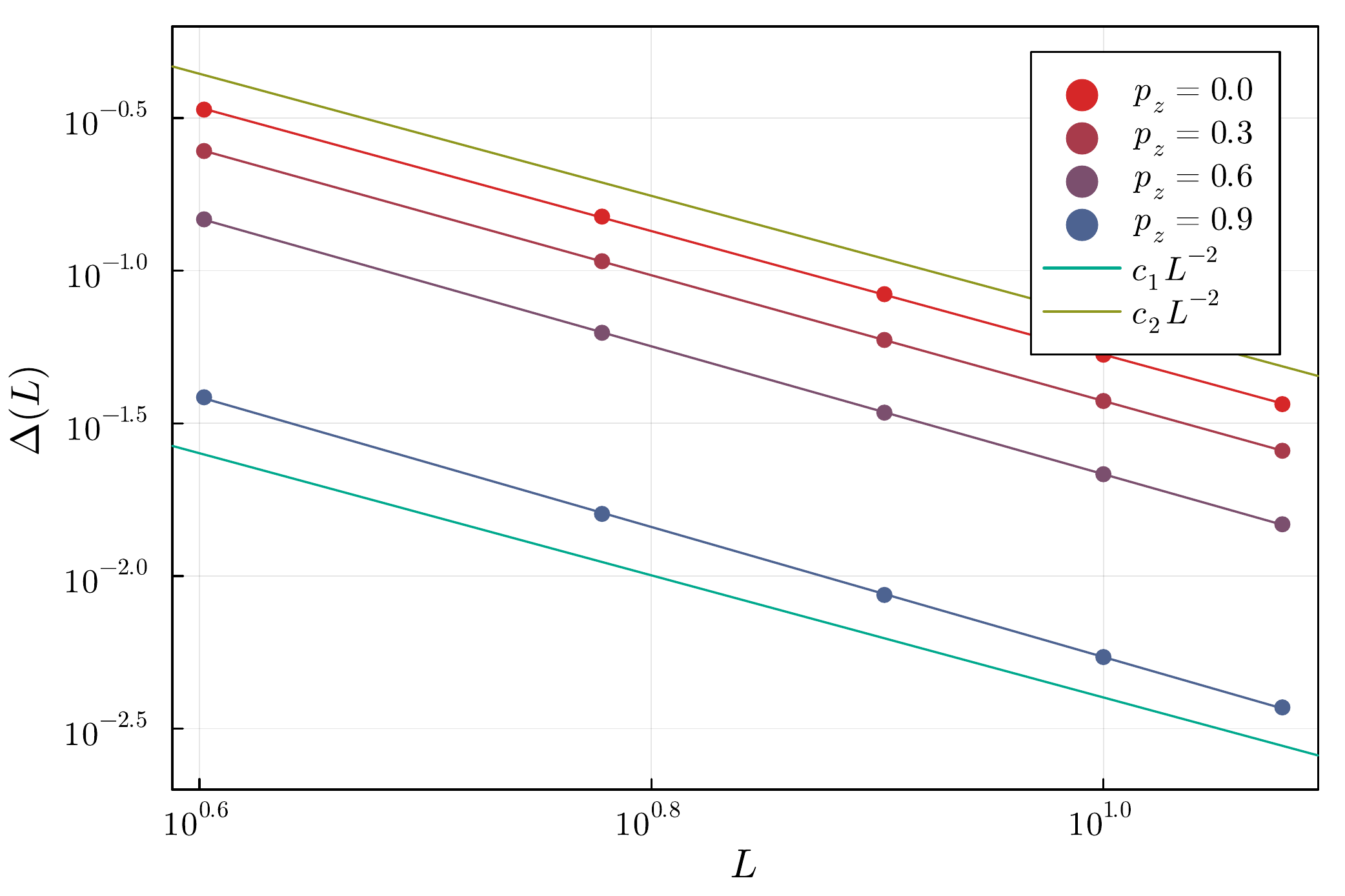}
    \caption{Channel gap scaling with system size for a range of $p_z$ values in the perturbed model with $U(1)$ symmetry (see Sec.~\ref{sec:results:U1_perturbations}). The data are arranged between guide lines $\Delta(L) = c_1L^{-2}$ and $\Delta(L) = c_2L^{-2}$ where $c_1 = 0.4$ and $c_2 = 7.0$. We consequently observe that gap scaling for each value of $p_z$ is consistent with diffusive dynamics.}
    \label{fig:U1_channel_gap}
\end{figure}

As in the baseline model, we can study the channel matrix gap to understand the saturation dynamics of linear observables. As in Sec.~\ref{sec:results:baseline_dynamics:approach}, we find gap scaling consistent with $\Delta \propto L^{-2}$ and diffusive dynamics. These results are plotted in Fig.~\ref{fig:U1_channel_gap}.

Despite this, we observe constant-time saturation of $\expval{\v{S}_1\cdot\v{S}_L}$ and entanglement entropy to their steady-state values ($\expval{\v{S}_1\cdot\v{S}_L} = 0$ and area-law entanglement in the thermodynamic limit). The numerical results are not shown here. But it is not surprising that the channel gap indicates $O(L^2)$ saturation time even when the quantities we have focused on saturate more quickly. The channel gap needs to account for the slowest dynamics of linear observables, and there are slower processes in these dynamics: for example, the diffusion of a localized charge requires $O(L^2)$ time steps.

\subsection{Perturbed dynamics without $U(1)$ symmetry}
\label{sec:results:general_perturbations} 

\subsubsection{Steady state properties}
\label{sec:results:general_perturbations:steady_state}
\begin{figure}[h!]
    \centering
    \includegraphics[width=\linewidth]{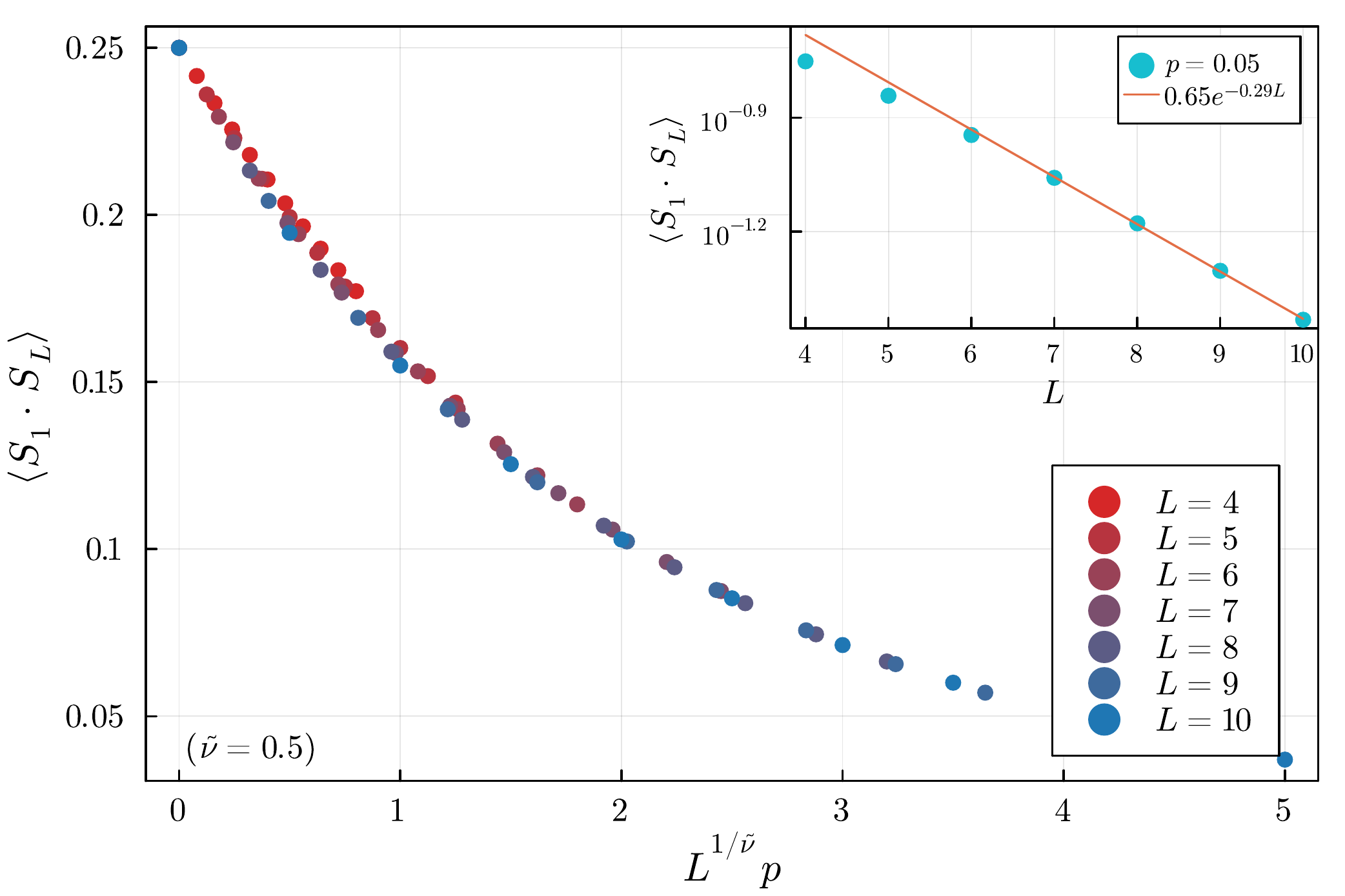}
    \caption{Steady state $\expval{\v{S}_1\cdot\v{S}_L}$ of the quantum channel (at half-filling) as a function of $L^{1/\tilde{\nu}}p$ in the perturbed model \emph{without} $U(1)$ symmetry (see Sec.~\ref{sec:results:general_perturbations}). Here, $p = p_x = p_y$ controls the symmetry-violating perturbation. We find that the data collapse as $\Tr[(\v{S}_1\cdot\v{S}_L)\rho(L,p)] = \tilde{F}_{SS}(L^{1/\tilde{\nu}}p)$ when $\tilde{\nu} \approx 0.5$. When $p > 0$, the steady state value of $\expval{\v{S}_1\cdot\v{S}_L}$ decays exponentially with system size, as illustrated for $p = 0.05$ in the inset. The linear fit in the inset results from a linear regression on the four data points with $L \geq 7$.}
    \label{fig:general_SS_scaling}
\end{figure}

If one or both of $p_x$ and $p_y$ is greater than zero, then the $U(1)$ symmetry is no longer present. In this case, we find that the steady states lack order and have area-law entanglement as long as multiple species of Pauli measurements are present. As in Sec.~\ref{sec:results:U1_perturbations:steady_state}, we can perform a finite-size scaling analysis to understand how order arises as we approach $p_s = 1$ from a direction in theory-space that does not have $U(1)$ symmetry. The results of this analysis are depicted in Fig.~\ref{fig:general_SS_scaling} where $p_s = 1$ is approached along the line $p = p_x = p_y$. We find that the data collapse to the scaling form
\[
\Tr[(\v{S}_1\cdot\v{S}_L)\rho(L,p)] = \tilde{F}_{SS}(L^{1/\tilde{\nu}}p)
\]
with $\tilde{\nu} \approx 0.5$.

\subsubsection{Approach to steady state}
\label{sec:results:general_perturbations:approach}
In the presence of symmetry-breaking perturbations, we find that the channel gap is generally independent of system size (data not shown). This is unsurprising: in the absence of $U(1)$ symmetry there are no longer conserved charges to diffuse throughout the system. Also, since the system only ever has area-law entanglement, entanglement saturation can occur in constant time.

\subsubsection{Exceptional cases}
\label{sec:results:general_perturbations:special}
A special situation arises when only one type of Pauli measurement is present. In Sec.~\ref{sec:results:U1_perturbations}, we found that two pure steady states arose in the perturbed dynamics with $U(1)$ symmetry: the states $\ket{{\uparrow}{\uparrow}\cdots{\uparrow}}$ and $\ket{{\downarrow}{\downarrow}\cdots {\downarrow}}$. However, charge conservation prevented these states from being accessible to initial states in non-extremal charge sectors, so most initial states led to mixed steady states.

When only $\sigma^x$ perturbations are present, the product state entirely pointed in $+x$ and the product state entirely pointed in $-x$ are both pure steady states (and likewise with states in $\pm y$ when only $\sigma^y$ perturbations are present). But, since there is no symmetry to enforce charge conservation of any kind, these pure steady states are generally accessible to the dynamics regardless of initial state. Consequently, when only $p_x$ is nonzero or only $p_y$ is nonzero, the steady state is pure and ferromagnetically ordered.

These steady states may correspond to channels with exceptional gap scaling (in comparison to the system-size-independent gap in Sec.~\ref{sec:results:general_perturbations:approach}). Even though there is no symmetry-breaking order being developed over time, we observe gapless modes in the channel matrix and slow saturation in the quantum trajectory data. Both indicate that the approach to these ordered steady states occurs on a time scale that grows with system size (data not shown). This is in stark contrast with the generic situation, when multiple species of Pauli measurements are present and the gapped channel matrix ensures that the steady state is quickly reached. We leave further exploration of these slow dynamics for future work.

\section{Discussion}\label{sec:discussion}
\subsection{Experimental protocol}
Our protocol is composed of single-site unitaries and measurements, and of local two-site SWAP measurements. These SWAP measurements can be implemented in a straightforward way using an ancilla qubit for each bond and Fredkin (CSWAP) gates controlled by these ancillae.

Suppose that our system is in state $\ket{\psi}$ and we wish to measure SWAP between sites $i$ and $i+1$. The projectors corresponding to the two possible outcomes are denoted $\Pi^{\pm}_{i,i+1}$. We can always write $\ket{\psi} = \ket{\psi^+}+\ket{\psi^-}$ where $\ket{\psi^{\pm}} = \Pi^{\pm}_{i,i+1}\ket{\psi}$ are unnormalized states and $\text{SWAP}_{i,i+1}\ket{\psi^{\pm}} = \pm\ket{\psi^{\pm}}$. It is convenient to denote the $x$-basis states as $\ket{\pm} = \frac{1}{\sqrt{2}}(\ket{{\uparrow}}\pm\ket{{\downarrow}})$.

Initially, the ancilla qubit for the $i$th bond is in the state $\ket{+}$ so the system and this ancilla together are in the state:
\[
\ket{+}\ket{\psi} = \tfrac{1}{\sqrt{2}}(\ket{{\uparrow}}+\ket{{\downarrow}})(\ket{\psi^+}+\ket{\psi^-}).
\]
Next, let $U$ denote a CSWAP gate that is controlled by the ancilla and that swaps the sites $i$ and $i+1$. When we apply $U$ to this system, we get:
\[
U\ket{+}\ket{\psi} &= \subalign{\tfrac{1}{\sqrt{2}}\big[&\ket{{\uparrow}}(\ket{\psi^+}+\ket{\psi^-})\\
&+\ket{{\downarrow}}(\ket{\psi^+}-\ket{\psi^-})\big]}\\
&= \ket{+}\ket{\psi^+} + \ket{-}\ket{\psi^-}.
\]
The SWAP measurement is performed on $\ket{\psi}$ when we measure the ancilla in the $x$ basis. We get the even-parity result with probability $\braket{\psi^+}{\psi^+} = \mel{\psi}{\Pi^+_{i,i+1}}{\psi}$ and the odd-parity result with probability $\braket{\psi^-}{\psi^-} = \mel{\psi}{\Pi^-_{i,i+1}}{\psi}$. These are the same probabilities one would find for a direct measurement of SWAP, so this process precisely implements a SWAP measurement. The ancilla can be reset and used for subsequent measurements across this bond.

\begin{figure}
    \centering
    \includegraphics[width=\linewidth]{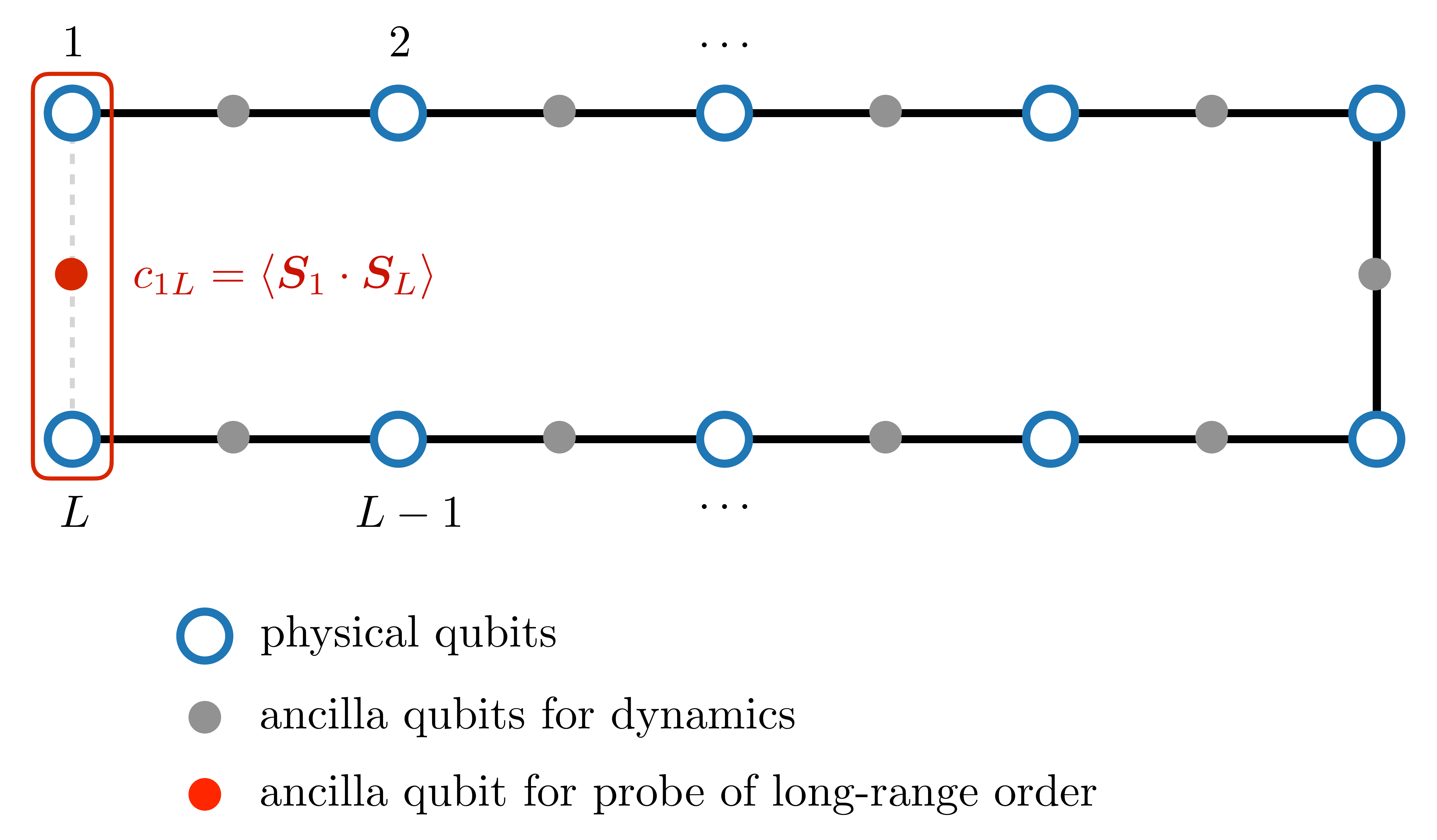}
    \caption{Schematic diagram depicting a simple implementation of the dynamics studied in this paper. Physical qubits are marked with blue circles and grey ancilla qubits lie along each bond. These ancillae enable local SWAP measurements. The red ancilla qubit lies between the first and last physical qubits. These physical qubits do not interact directly in our models (which have open boundary conditions) but a SWAP measurement between them diagnoses long-range order.}
    \label{fig:experimental}
\end{figure}

Since the feedback in our baseline model produces long-range order that is linear in the density matrix, a decoherence-free quantum computer could observe this order without a postselection problem. We found that this order can be probed by $\mel{\psi}{\v{S}_1\cdot\v{S}_L}{\psi}$. Since $\v{S}_1\cdot\v{S}_L = \frac{1}{4}(2\text{SWAP}_{1L}-1)$, this is equivalent to measuring SWAP between the first and last sites of the system. If the $L$ qubits comprising our system are arranged in a loop, then this long-range order is locally accessible experimentally using an ancilla between the first and last sites. Such a setup is depicted in Fig.~\ref{fig:experimental}.

\subsection{Outlook}
In this work, we explored how adaptive quantum dynamics can be used to target pure steady states that break continuous symmetries and host long-range entanglement. In particular, we introduced a baseline model where the steady state breaks a continuous $U(1)$ symmetry and has logarithmic entanglement. But instead of finding a stable phase with these properties, we determined that any nonzero density of Pauli measurements leads to a mixed steady state without order and with area-law entanglement in the quantum trajectories. 


This prompts the question: is it possible to have a robust \emph{phase} that spontaneously breaks a continuous symmetry in adaptive quantum circuits? This could arise from a stable absorbing phase with a symmetry-breaking pure steady state. Alternatively, perhaps a mixed steady state could arise that breaks a continuous symmetry. Whether such an absorbing phase or such a mixed state are possible could be interesting future directions.

Also, it is notable that the symmetry-breaking steady state of our baseline model is also a ground state of a frustration-free Hamiltonian (the ferromagnetic Heisenberg model). And this is recapitulated in the dynamics we consider: the steady state is simultaneously stabilized by SWAP measurements with feedback at every bond. This is an appealing feature of the dynamics because it ensures that, at least in the absence of perturbations, there exist pure steady states. In contrast, there is no non-trivial state that is stabilized by a measurement like $X_iX_{i+1}+Y_iY_{i+1}$ at all bonds $(i,i+1)$. It would be interesting to understand whether or not there is a fundamental obstruction to stabilizing ground states of quantum Hamiltonians for which the ground state does not simultaneously minimize each term (like the quantum XY model).

Lastly, we identified a non-Hermitian ``Hamiltonian'' whose ground states encode the steady state properties of a quantum channel and whose gap determines how quickly the steady state is reached. Since the quantum channel has full information about linear observables, this amounts to the quantities that can be generally observed experimentally. Consequently, it could be interesting to explore whether properties of these non-Hermitian operators can be used to organize experimentally-accessible properties of quantum channels.

\section{Acknowledgements}
We acknowledge helpful discussions with Tim Hsieh, Ali Lavasani, Roger Melko, and Giacomo Torlai.

This material is based upon work supported by the National Science Foundation Graduate Research Fellowship Program under Grant No.~2139319 (J.H.), by the Simons Collaboration on Ultra-Quantum Matter, which is a grant from the Simons Foundation (651457, M.P.A.F. and J.H.).
Y.L. is supported in part by the Gordon and Betty Moore Foundation’s EPiQS Initiative through Grant GBMF8686, and in part by the Stanford Q-FARM Bloch Postdoctoral Fellowship.
This work is supported in part by the National Science Foundation under Grant No.~NSF PHY-1748958. Use was made of computational facilities purchased with funds from the National Science Foundation (CNS-1725797) and administered by the Center for Scientific Computing (CSC). The CSC is supported by the California NanoSystems Institute and the Materials Research Science and Engineering Center (MRSEC; NSF DMR 1720256) at UC Santa Barbara.

\bibliography{refs}

\clearpage

\appendix

\section{The doubled Hilbert space}\label{appendix:doubled_hilbert_space}
The key elements of translating expressions in the standard Hilbert space to a doubled Hilbert space are contained in the following rearrangement. Given a basis $\{\ket{b_i}\}_{i=1}^{2^L}$, we can write
\[
\rho &= \sum_{ij} \rho_{ij} \dyad{b_i}{b_j}\text{,}\\
K &= \sum_{ij} K_{ij} \dyad{b_i}{b_j}\text{, and}\\
K^\dag &= \sum_{ij} (K^\dag)_{ij} \dyad{b_i}{b_j}\text{,}
\]
and notice that
\[
(K\rho K^\dag)_{ij} &= \sum_{mn} K_{im}\rho_{mn} (K^\dag)_{nj}\\
&= \sum_{mn} K_{im}(K^*)_{jn}\rho_{mn} \\
&= \braa{b_ib_j}(K\ot K^*)\kett{\rho},
\]
where
\[
\kett{\rho} &= \sum_{ij} \rho_{ij} \kett{b_ib_j}
\]
with $\kett{b_ib_j} = \ket{b_i}\ot\ket{b_j}$ and $\braa{b_ib_j} = \bra{b_i}\ot\bra{b_j}$.

But there is a noteworthy subtlety: in doing this transformation, we took a fixed basis (the computational basis), so the process of vectorizing $\rho$ and transposing $K^\dag$ is basis-dependent.

In the doubled Hilbert space, we notice that
\[
\Tr(A^\dag B) &= \sum_{ij} (A^\dag)_{ji} B_{ij}\\
&= \sum_{ij} A^*_{ij} B_{ij}\\
&= \braakett{A}{B}.
\]
Also,
\[
\braakett{b_ib_j}{\rho^\dag} &= (\rho^\dag)_{ij}\\
&= \rho_{ji}^*\\
&= \braakett{b_jb_i}{\rho}^*
\]
so
\[\label{eq:doubled_hermitian}
\kett{\rho^\dag} = \left(T\kett{\rho}\right)^*
\]
where $T$ swaps bitstrings between the two copies of the Hilbert space. When $\rho$ is a physical Hermitian state, Eq.~\ref{eq:doubled_hermitian} provides a condition on the vectorized form of the state.

\section{Channel gaps}\label{appendix:channel_gaps}
A monitored quantum circuit's saturation time scaling is particularly transparent in the quantum channel picture. The entire spectrum of any quantum channel $\mathcal{E}$ is contained in the unit disk of $\mathbb{C}$~\cite{QChannelLecture}. If $\{V_k\}$ is a (generally non-orthonormal) eigenbasis for $\mathcal{E}$ with corresponding eigenvalues $\{\lambda_k\}$, then any initial state $\rho_0$ can be written as
\[
\rho_0 = \sum_k a_k V_k,
\]
and after $M$ time steps,
\[
\mathcal{E}^M(\rho_0) = \sum_k \lambda_k^M a_k V_k.
\]
One or more $\lambda_k$ will satisfy $\abs{\lambda_k} = 1$, so that as $M \to \infty$ the steady state is given by
\[
\mathcal{E}^M(\rho_0) = \sum_{k:\abs{\lambda_k}=1} \lambda_k^M a_k V_k
\]
and the remaining unsteady contributions to the state will shrink as
\[
\sum_{k:\abs{\lambda_k}<1} \lambda_k^M a_k V_k &= \abs{\lambda}^M \sum_{k:\abs{\lambda_k}<1} \left(\frac{\lambda_k}{\abs{\lambda}}\right)^M a_k V_k
\]
where $\lambda$ is the largest eigenvalue satisfying $\abs{\lambda} < 1$. Since $\abs{\lambda} < 1$, there is a gap $\Delta$ between $\abs{\lambda}$ and the largest eigenvalue $1$. Generically, this gap could depend on system size, so that the unsteady parts of the initial state fall off with
\[
\abs{\lambda}^M = (1-\Delta(L))^M,
\]
meaning that $M = O(\Delta(L)^{-1})$ time steps are required to ensure that these unsteady parts become arbitrarily small.

In other words, the saturation time for a channel $\mathcal{E}$ scales with the inverse of the gap between $1$ and the next largest eigenvalue of $\mathcal{E}$, or with the inverse of the gap in the non-Hermitian ``Hamiltonian''
\[
H = \one - \mathcal{E}.
\]
As noted in Sec.~\ref{sec:results:baseline_dynamics:approach}, there exist exceptions to this analysis (like that explored in the context of Lindblad evolution in~\cite{Mori_2020}) that arise from the non-Hermiticity of~$H$.

\section{$Z_iZ_j$ correlations in perturbed dynamics with $U(1)$ symmetry \label{appendix:ZZ}}
In this Appendix \ref{appendix:ZZ}, we calculate $\Tr[S_i^zS_j^z \rho(L,p_z)]$ for $i\neq j$, where $\rho(L,p_z)$ is the steady state of the perturbed dynamics with $U(1)$ symmetry in a given charge sector. We will first show that this expectation value is independent of $i$ and $j$, and then we will use this to derive the value for all $i \neq j$.

Using the doubled Hilbert space formalism of Appendix \ref{appendix:doubled_hilbert_space}, we can write
\[
\Tr[S_i^zS_j^z \rho(L,p_z)] = \braakett{S_i^zS_j^z}{\rho(L,p_z)}
\]
where, taking $\{\ket{b_i}\}_{i=1}^{2^L}$ to be the set of computational basis states,
\[
\braa{S_i^zS_j^z} &= \sum_{m,n} \braa{b_mb_n} (S_i^zS_j^z)^*_{mn}\\
&= \sum_{m} \braa{b_mb_m}\mel{b_m}{S_i^zS_j^z}{b_m}
\]
since $S_i^zS_j^z$ is diagonal in the computational basis. 

When we set $p_x = p_y = 0$, the perturbed channel is given by
\[
\C' &= \subalign{\frac{1}{L-1}&\sum_{i=1}^{L-1}p_s\left[(\Pi^{+}_{i,i+1})^{\ot 2}+(\sigma^z_i\Pi^{-}_{i,i+1})^{\ot 2} \right]\\
+\frac{1}{2L}&\sum_{i=1}^Lp_z\left[\one^{\ot2}+(\sigma_i^z)^{\ot2}\right].}
\]
Defining $\Pi^{Z\ot Z \pm}_i = (1 \pm (\sigma_i^z)^{\ot2})/2$, we can rewrite this channel as
\[
\C' &= \subalign{\frac{1}{2}\frac{1}{L-1}&\sum_{i=1}^{L-1}\subalign{p_s\big[ &\Pi^{Z\ot Z +}_i\left(\one^{\ot2}+ \text{SWAP}_{i,i+1}^{\ot2}\right)\\
+\, &\Pi^{Z\ot Z -}_i\subalign{&\big(\text{SWAP}_{i,i+1}\ot\one\\
&+ \one\ot\text{SWAP}_{i,i+1}\big)\big]}}\\
+\frac{1}{L}&\sum_{i=1}^Lp_z\Pi^{Z\ot Z +}_i.}
\]
Given an initial state $\kett{\rho_0}$ with charge $Q$, the repeated application of $\C'$ will yield the steady state of the channel in that charge sector:
\[
\kett{\rho(L,p_z)} = \lim_{T \to \infty} (\C')^T\kett{\rho_0}.
\]
Therefore,
\[
\braakett{S_i^zS_j^z}{\rho(L,p_z)} = \lim_{T \to \infty}\braa{S_i^zS_j^z}(\C')^T\kett{\rho_0}.
\]
To calculate this, we can consider the action of $\C'$ to the left on $\braa{S_i^zS_j^z}$. This is vastly simplified by noticing that $\braa{b_mb_m} (\sigma_i^z)^{\ot 2} =\braa{b_mb_m} $ for all computational basis states $\ket{b_m}$, so
\[
\braa{b_mb_m} \Pi^{Z\ot Z +}_i &= \braa{b_mb_m} 
\]
and
\[
\braa{b_mb_m} \Pi^{Z\ot Z -}_i &= 0.
\]
Consequently,
\[
\braa{b_mb_m}  \C' = \braa{b_mb_m}  \tilde{\C}'
\]
for all $b_m$, where
\[
\tilde{\C}' = \frac{1+p_z}{2} + \frac{1-p_z}{2}\frac{1}{L-1}\sum_{i=1}^{L-1} \text{SWAP}_{i,i+1}^{\ot 2}.\label{eq:projected_channel}
\]
Also, since $\braa{b_mb_m} \text{SWAP}_{i,i+1}^{\ot 2} = \braa{b_nb_n} $ for some $n$ (where $b_n$ is obtained from $b_m$ by swapping its $i$th and $(i+1)$th bits), it follows that $\braa{b_mb_m}$ remains in the subspace spanned by $\{\braa{b_ib_i}\}_{i=1}^{2^L}$ when multiplied by $\C'$. Therefore, the same simplification holds for repeated application of $\C'$, so
\[
\braa{b_mb_m}  (\C')^T = \braa{b_mb_m} (\tilde{\C}')^T
\]
and, consequently, 
\[
\braa{S_i^zS_j^z} (\C')^T = \braa{S_i^zS_j^z} (\tilde{\C}')^T.
\]
The matrix $\tilde{\C}'$ is Hermitian (even though the original channel matrix ${\C}'$ was not) so its left eigenvectors and right eigenvectors are equal. Furthermore, it is the matrix corresponding to a quantum channel,\footnote{More specifically, the channel matrix in Eq.~\ref{eq:projected_channel} corresponds to a model where SWAP is measured at a random bond with probability $1-p_z$ and nothing occurs with probability $p_z$.} so it must have at least one steady state.

In fact, in the subspace spanned by $\{\braa{b_ib_i}\}_{i=1}^{2^L}$, there is exactly one steady state of $\tilde{\mathcal{C}}'$ in each charge sector. This steady state should be an eigenvector of $\text{SWAP}_{i,i+1}^{\ot 2}$ with eigenvalue $+1$ for all $i$. Since $\bra{b_n}\text{SWAP}_{ij}\ket{b_m}$ is always $0$ or $1$, it follows that
\[
\braa{b_nb_n}\text{SWAP}_{ij}^{\ot 2}\kett{b_mb_m} &= \bra{b_n}\text{SWAP}_{ij}\ket{b_m}^2\\
&= \bra{b_n}\text{SWAP}_{ij}\ket{b_m}.
\]
Therefore, $\text{SWAP}_{ij}^{\ot 2}$ has the same matrix elements in the space spanned by $\{\braa{b_ib_i}\}_{i=1}^{2^L}$ as $\text{SWAP}_{ij}$ does in the original single-copy Hilbert space, of which the space spanned by $\{\braa{b_ib_i}\}_{i=1}^{2^L}$ is a repetition code. Considering the single-copy Hilbert space, we saw in Sec.~\ref{sec:results:baseline_dynamics:steady_state} that only states with maximal total spin are even under $\text{SWAP}_{i,i+1}$ for all $i$. It follows that in the repetition code, the steady states are
\[
\braa{\tfrac{L}{2},Q} \propto \sum_{m : w(b_m)=N} \braa{b_mb_m}\label{eq:repeated_steady_states}
\]
where $N = Q+\frac{L}{2}$ is the total number of spins pointing up (in the $+z$ direction) and $w(b_m)$ counts the number of up spins in the computational basis state $\ket{b_m}$. It follows that
\[
\braakett{S_i^zS_j^z}{\rho(L,p_z)} &= \lim_{T \to \infty}\braa{S_i^zS_j^z}(\C')^T\kett{\rho_0}\\
&= \lim_{T \to \infty}\braa{S_i^zS_j^z}(\tilde{\C'})^T\kett{\rho_0}\\
&\propto \braakett{\tfrac{L}{2},Q}{\rho_0}.
\]
As a result, we conclude that this expectation value is independent of $i$ and $j$.

Now we are ready to finish our calculation. Since $\left(\sum_{i=1}^L S_i^z\right)\rho(L,p_z) = Q\rho(L,p_z)$,
it follows that
\[
Q^2 &= \Tr[\left(\sum_{i=1}^L S_i^z\right)^2\rho(L,p_z)]\\
&= \frac{L}{4} + L(L-1)\Tr[S_i^z S_j^z \rho(L,p_z)]
\]
using our result that $\Tr[S_i^z S_j^z \rho(L,p_z)]$ is independent of $i$ and $j$. Rearranging this, we find that
\[
\Tr[S_i^z S_j^z \rho(L,p_z)] = \frac{4Q^2-L}{4L(L-1)}.
\]
We notice that this result is independent of $p_z$.

\section{Steady state entanglement \label{appendix:entanglement}}
As stated in Sec.~\ref{sec:results:baseline_dynamics:steady_state}, we find that the steady state of the baseline model has logarithmic entanglement when the initial state lies in a single charge sector with $Q = qL$ and $q$ satisfying $-\frac{1}{2} < q < \frac{1}{2}$. When $Q$ is held fixed with respect to system size, the steady state has area-law entanglement. In this Appendix \ref{appendix:entanglement}, we derive these results for bipartitions of the system into regions $A$ and $B$, where $A$ contains the first $|A|$ spins and $B$ contains the remaining $|B|$ spins with $|A|+|B| = L$.

\subsection{Determining the reduced density matrix\label{appendix:entanglement:density_matrix}}
Let $N = Q + \frac{L}{2}$ be the number of spins pointing up (in the $z$ direction). Then the steady state is
\[
\ket{\psi} = \frac{1}{\sqrt{\mathcal{N}}} \sum_{b:w(b)=N}\ket{b},\label{eq:steady_state}
\]
where the sum is over bitstrings denoting up and down spins, with $w(b)$ counting the number of up spins. There are $\binom{L}{N}$ such bitstrings, so $\mathcal{N} = \binom{L}{N}$. Since the state's entanglement is unchanged if we flip each bit, we need only consider $N \leq \tfrac{L}{2}$. Focusing on the left part of the system, we have:
\[
\mel{b_A}{\rho_A}{b_A'} &= \sum_{b''} \mel{b_A b''_{B}}{\rho}{b'_A b''_{B}}\\
&= \frac{1}{\mathcal{N}}\sum_{\substack{c,c'\\w(c)=w(c')=N}} \sum_{b''} \braket{b_A b''_{B}}{c}\braket{c'}{b'_A b''_{B}}\\
&= \frac{1}{\mathcal{N}}\sum_{\substack{c_A,c_A'\\w(c_A)=w(c_A')}}\binom{|B|}{N-w(c_A)}\braket{b_A}{c_A}\braket{c_A'}{b'_A}\\
&= \delta_{w(b_A),w(b_A')}\frac{\binom{|B|}{N-K}}{\binom{L}{N}}
\]
where $K = w(b_A)$. The final line follows since $w(c_A)=w(c_A')$ forces $w(b_A)=w(b_A')$ for nonzero terms. It follows that
\[
\rho_A &= \bigoplus_{K=0}^{\min(|A|,N)} \rho_A^{(K)}\\
&= \bigoplus_{K=0}^{\min(|A|,N)} \frac{\binom{|B|}{N-K}}{\binom{L}{M}} \pmat{1 & \hdots & 1\\
\vdots & \ddots & \vdots \\
1 & \hdots & 1}_{\binom{|A|}{K}\times\binom{|A|}{K}}.
\]
The sum over sectors is from the minimum possible number of up spins in $A$ (which is $\max(0,N-|B|)$) to the maximum possible number ($\min(|A|,N)$). But since we can calculate the entanglement entropy equivalently from $\rho_A$ or $\rho_B$, we can pick $|A| \leq \frac{L}{2}$ and $|B| \geq \frac{L}{2}$ which, having already chosen $N \leq \frac{L}{2}$, reduces the lower bound to $0$ without loss of generality.

The $n\times n$ all-ones matrix has spectrum $\{n,0,\dots,0\}$ so the nonzero elements of the spectrum of $\rho_A$ are
\[
p_K = \frac{\binom{|A|}{K}\binom{|B|}{N-K}}{\binom{L}{N}}
\]
for $K$ from $0$ to $\min(|A|,N)$. It follows that
\[
S(\rho_A) = -\sum_{K=0}^{\min(|A|,N)} p_K \log p_K.\label{eq:entanglement_sum}
\]

\subsection{Simplifying and bounding the entanglement\label{appendix:entanglement:bounding}}
If we define intensive quantities such that $|A| = aL$, $|B| = bL$, $N = nL$, and $K = kL$, and define $\Delta k = \frac{1}{L}$, then we arrive at
\[
S(\rho_A) &= -L\sum_{K=0}^{L\min(a,n)}(\Delta k)\, p_k \log p_k\label{eq:entanglement_sum_2}
\]
where
\[
p_k = \frac{\binom{aL}{kL}\binom{bL}{(n-k)L}}{\binom{L}{nL}}.
\]
According to Stirling's formula, the following asymptotic relation holds:
\[
x! \sim \sqrt{2\pi x}\left(\frac{x}{e}\right)^x .
\]
Applying this formula to the combinatorial factors in $p_k$ yields, after a straightforward calculation:
\[
\log p_k \sim f(k)L - \tfrac{1}{2}\log2\pi L + \tfrac{1}{2}g(k)  + O(L^{-1})
\]
where
\[
f(k) &= \subalign{&a\log a + b\log b\\
&+ n\log n+(1-n)\log(1-n)\\
&- k\log k-(a-k)\log(a-k)\\
&- (n-k)\log(n-k)-(b-n+k)\log(b-n+k)}
\]
and
\[
g(k) &= \subalign{&\log a + \log b + \log n+\log(1-n)\\
&- \log k - \log(a-k) - \log(n-k)- \log(b-n+k).}
\]
If we define the additional shorthand
\[
h(k) = e^{\frac{1}{2}g(k)}\left(f(k)L-\tfrac{1}{2}\log 2\pi L +\tfrac{1}{2}g(k) \right)
\]
then
\[
p_k \log p_k \sim \sqrt{\frac{1}{2\pi L}}e^{Lf(k)}h(k). \label{eq:plogp}
\]
We note that $f(k)$ is always non-positive, which prevents Eq.~\ref{eq:plogp} from diverging exponentially with $L$. We verify this below when we find (in Eq.~\ref{eq:f0}) that the maximum value of $f(k)$ is $0$.

In what follows, we seek to apply this approximation to Eq.~\ref{eq:entanglement_sum_2} and to approximate that sum as an integral, in order to analytically extract leading order behaviour in $L$. Doing this rigorously requires some care: we will sandwich the entanglement between two Riemann sums related to Stirling's approximation, and show that each converges to a well-behaved integral in the thermodynamic limit.

\begin{figure}[h]
    \centering
    \includegraphics[width=\linewidth]{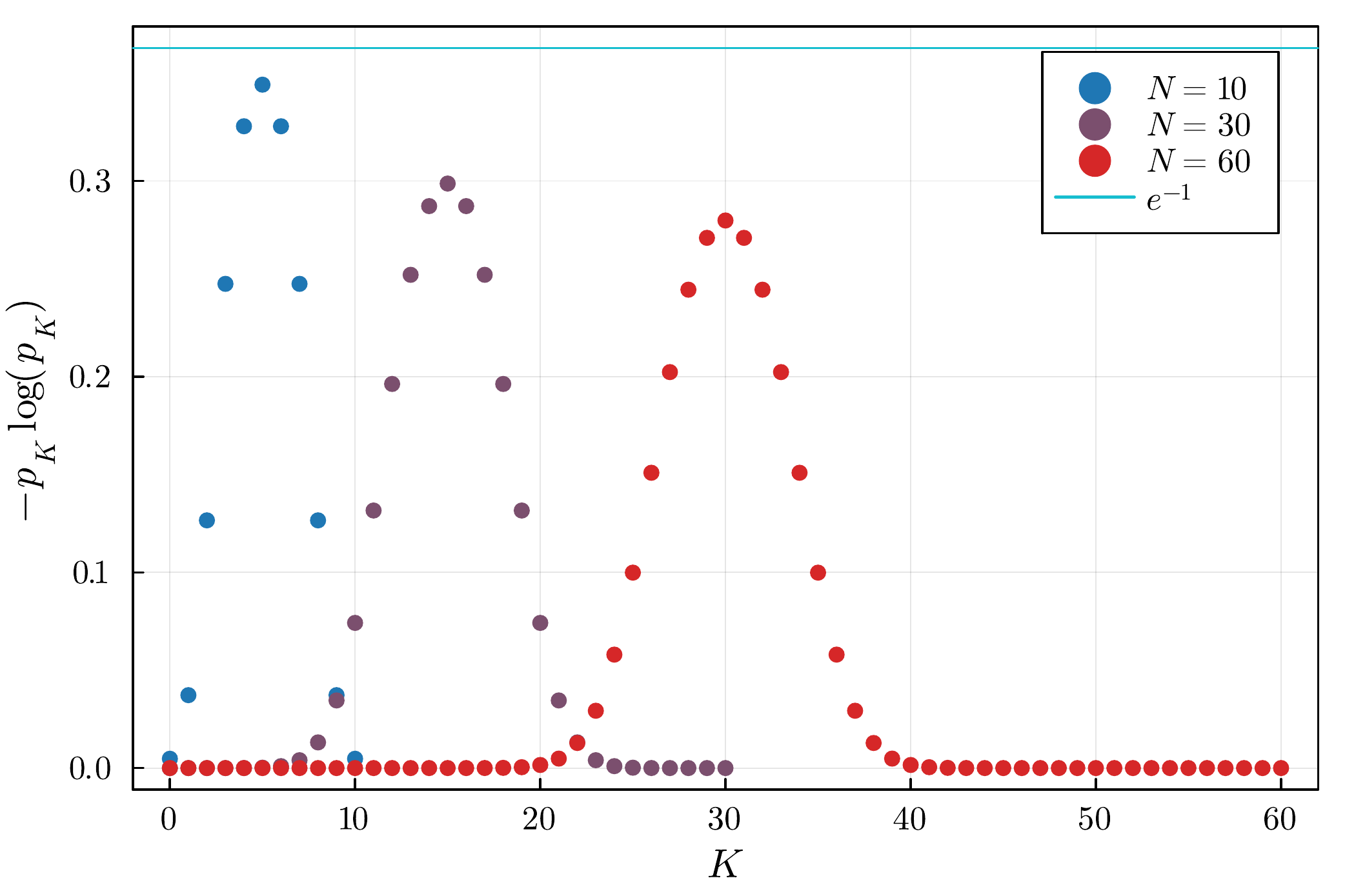}
    \caption{Values of $-p_K\log p_K$ (summands in Eq.~\ref{eq:entanglement_sum}) for a system with $L=100$, partitioned in two equal halves ($|A| = |B| = L/2$), over a range of fillings. The maximum value of this function, which is $e^{-1}$, is plotted for reference. It turns out that $p_K$ is invariant under $\abs{A} \leftrightarrow N$, so this plot can also be viewed as depicting $-p_K\log p_K$ for a system at half-filling over a range of partition sizes.}
    \label{fig:appendix_integrand}
\end{figure}

But it is not surprising that the integrals we encounter converge and are well-behaved. The entanglement density $\sigma(\rho_A) = S(\rho_A)/L$ must satisfy $0 \leq \sigma(\rho_A) \leq  a\log 2$ because the entanglement is non-negative and the maximally-mixed reduced density matrix on $|A|$ sites has entropy $|A|\log 2$. Furthermore, $-p\log p$ always lies within $[0,e^{-1}]$ for $p \in [0,1]$, so when we view Eq.~\ref{eq:entanglement_sum_2} as a Riemann sum, its integrand is always well-behaved. Values of this integrand are depicted in Fig.~\ref{fig:appendix_integrand} for a range of fillings (or, as is explained in the caption, for a range of partition sizes).

There is a statement related to Stirling's approximation which holds that
\[
\sqrt{2\pi x}\left(\frac{x}{e}\right)^x e^{\frac{1}{12x+1}} < x! <\sqrt{2\pi x}\left(\frac{x}{e}\right)^x e^{\frac{1}{12x}}
\]
is true for all $x \geq 1$~\cite{Robbins_1955}. Consequently,
\[
\sqrt{\frac{1}{2\pi L}} e^{f(k)L+\frac{1}{2}g(k)+r_1(k)} < p_k < \sqrt{\frac{1}{2\pi L}} e^{f(k)L+\frac{1}{2}g(k)+r_2(k)}
\]
where
\begin{widetext}
\[
r_1(k) &= \frac{1}{1+12aL}+\frac{1}{1+12bL} +\frac{1}{1+12nL}+\frac{1}{1+12(1-n)L} -\frac{1}{12L}\left(\frac{1}{k}+\frac{1}{a-k}+\frac{1}{n-k}+\frac{1}{b-n+k}+1 \right)\\
r_2(k) &= \frac{1}{12L}\left(\frac{1}{a}+\frac{1}{b}+\frac{1}{n}+\frac{1}{1-n} \right)-\bigg(\frac{1}{1+12kL}+\frac{1}{1+12(a-k)L}+\frac{1}{1+12(n-k)L}+\frac{1}{1+12(b-n+k)L}+\frac{1}{1+12L}\bigg)
\] 
\end{widetext}
so
\[
p_k\log p_k &> \sqrt{\frac{1}{2\pi L}}e^{Lf(k)}e^{r_1(k)}\left(h(k)+e^{\frac{1}{2}g(k)}r_1(k)\right),\\
p_k\log p_k &< \sqrt{\frac{1}{2\pi L}}e^{Lf(k)}e^{r_2(k)}\left(h(k)+e^{\frac{1}{2}g(k)}r_2(k)\right).
\]

We further define
\[
\tilde{h}_i(k) &= h(k)+e^{\frac{1}{2}g(k)}r_i(k), \\
S_i(\rho_A) &= -\sqrt{\frac{L}{2\pi}}\sum_{K=0}^{L\min(a,n)}(\Delta k)e^{f(k)L}\tilde{h}_i(k)e^{r_i(k)},
\]
so that,
\[
S_1(\rho_A) < S(\rho_A) < S_2(\rho_A).
\]
For fixed $L$, each $S_i(\rho_A)$ is a Riemann sum, whose distance from the corresponding integral is given by:
\[
\Bigg|-\sqrt{\frac{L}{2\pi}}\int\displaylimits_{k=0}^{\min(a,n)}\D{k}e^{f(k)L}&\tilde{h}_i(k)e^{r_i(k)} - S_i(\rho_A)\Bigg| \\
&\leq \frac{M_i\min(a,n)}{2L}\\
\]
where $M_i$ is the maximum value of 
\[
\dd{}{k}(e^{f(k)L}\tilde{h}_i(k)e^{r_i(k)})
\]
over the interval. If we can show that $\lim_{L\to \infty} \frac{M_i}{L} = 0$ then we will have proven that the sums for $S_i(\rho_A)$ can be replaced by integrals. To do this, we first note that
\[
\dd{}{k}(e^{f(k)L}\tilde{h}_i(k)&e^{r_i(k)}) = e^{f(k)L}\big(Lf'(k)\tilde{h}_i(k)e^{r_i(k)}\\
&+\tilde{h}_i'(k)e^{r_i(k)}+\tilde{h}_i(k)r_i'(k)e^{r_i(k)}\big).
\]
Focusing on $f(k)$, we find that it is maximized at $k_* = an$, which always lies in the interval because $a,n < 1$, so $an < \min(a,n)$. At this maximum, we find that $f(k_*)=0$. Therefore, if $k \neq k_*$ then $M_i$ will have a negative exponential in $L$ that will dominate, for large $L$, over the subexponential orders in $L$ coming from the other factors. We conclude that $M_i$ must occur at $k=k_*$. It will be useful for what follows to find the leading order values of several functions at this point. First, we have:
\[
f(k_*) &= 0,\label{eq:f0}\\
f'(k_*) &= 0,\label{eq:f1}\\
f''(k_*) &= -[abn(1-n)]^{-1},\label{eq:f2}
\]
and
\[
g(k_*) &= -\log(abn(1-n)),\label{eq:g0}\\
g'(k_*) &= \frac{(2n-1)(b-a)}{abn(1-n)},\label{eq:g1}\\
g''(k_*) &= \frac{(1-2ab)(1-2n(1-n)}{[abn(1-n)]^2}.\label{eq:g2}
\]
Also, $r_i(k_*)$, $r_i'(k_*)$, and $r_i''(k_*)$ are all $O(L^{-1})$. Building the more complicated functions from these, we find that
\[
h(k_*) &= \frac{1}{\sqrt{abn(1-n)}}\left(-\tfrac{1}{2}\log 2\pi L  -\tfrac{1}{2}\log(abn(1-n)) \right)\\
&= -\frac{\log L}{2\sqrt{abn(1-n)}} -\frac{\log (2\pi abn(1-n))}{2\sqrt{abn(1-n)}},\label{eq:h0}\\
h'(k_*) &= \tfrac{1}{2}g'(k_*)(h(k_*)+e^{\tfrac{1}{2}g(k_*)})\\
&= \frac{(2n-1)(2a-1)\log L}{4(abn(1-n))^{3/2}} + O(1)\label{eq:h1}\\
h''(k_*) &= \subalign{&e^{\tfrac{1}{2}g(k_*)}\left[f''(k_*)L-\tfrac{1}{2}g''(k_*)-\tfrac{1}{4}g'(k_*)^2 \right]\\
&+\tfrac{1}{2}g''(k_*)h(k_*) + \tfrac{1}{2}g'(k_*)h'(k_*)}\\
&= -\frac{L}{(abn(1-n))^{3/2}} + O(\log L)\label{eq:h2}
\]
and that $\tilde{h}_i(k_*)$, $\tilde{h}_i'(k_*)$, and $\tilde{h}_i''(k_*)$ match these values at the orders given in Eqs.~\ref{eq:h0}, \ref{eq:h1}, and \ref{eq:h2}.

Having laid this groundwork, we conclude that
\[
M_i &= \subalign{\big[h'&(k_*)+e^{\frac{1}{2}g(k_*)}(\tfrac{1}{2}g'(k_*)r_i(k_*)+r_i'(k_*)) \\
&+ (h(k_*)+e^{\frac{1}{2}g(k_*)}r_i(k_*))r_i'(k_*)\big]e^{r_i(k_*)}}\\
&= \frac{(2n-1)(2a-1)\log L}{4(abn(1-n))^{3/2}} + O(1).
\]
It follows that
\[
\lim_{L\to \infty} \frac{M_i\min(a,n)}{2L} = 0,
\]
so
\[
S_i(\rho_A) = -\sqrt{\frac{L}{2\pi}}\int\displaylimits_{k=0}^{\min(a,n)}\D{k}e^{f(k)L}\tilde{h}_i(k)e^{r_i(k)}.
\]
Using Laplace's method, we will show that $S_1(\rho_A) = S_2(\rho_A)$ to leading order and next-to-leading order, allowing us to uniquely determine $S(\rho_A)$ at these orders.

\subsection{Laplace's method: leading order\label{appendix:entanglement:laplace1}}
In the thermodynamic limit, the integral is dominated by the neighbourhood of $k = k_*$, allowing us to approximate the integral using Laplace's method. We change variables to $x = k - k_*$ and notice that the integral bounds are irrelevant at leading order as ${L \to \infty}$ as long as $x = 0$ remains within them. Then, to leading order, Laplace's method yields:
\[
S_i(\rho_A) &\sim -\sqrt{\frac{L}{2\pi}} e^{Lf(k_*)} \tilde{h}_i(k_*)e^{r_i(k_*)}\int \D{x}e^{L\left(\frac{1}{2}f''(k_*)x^2\right)}\label{eq:laplace1}\\
&\sim \sqrt{\frac{-1}{f''(k_*)}}e^{Lf(k_*)}\tilde{h}_i(k_*)e^{r_i(k_*)}\label{eq:laplace3}.
\]
Applying Eqs.~\ref{eq:f0}, \ref{eq:f2}, and \ref{eq:h0}, we find that:
\[
S_i(\rho_A) = \tfrac{1}{2}\log L + O(1).
\]
Consequently, at leading order $S_1(\rho_A) = S_2(\rho_A)$, so we conclude that
\[
S(\rho_A) = \tfrac{1}{2}\log L + O(1).
\]
This demonstrates the logarithmic scaling of entanglement entropy for fixed non-extremal charge density.

When $N = nL$ is held fixed instead of $n$ in the thermodynamic limit, this does not apply. In this case, we can return to the original sum form of the entanglement entropy given in Eq.~\ref{eq:entanglement_sum}. Through our analysis in Sec.~\ref{appendix:entanglement:bounding}, we learned that $p_k \log p_k \sim \sqrt{\frac{1}{2\pi L}} e^{Lf(k)}h(k)$ is maximized at $k_* = an$. There, it takes the value
\[
p_k \log p_k \sim -\frac{\log(2\pi abN)}{2\sqrt{2\pi abN}} + O(L^{-1})
\]
when $N$ is held constant in the thermodynamic limit. Since $p_K$ and $p_k$ describe the same quantity, just in terms of a different variable, it follows that
\[
S(\rho_A) &= -\sum_{K=0}^{\min(|A|,N)} p_K \log p_K\\
&\leq -N \max(p_K \log p_K)\\
&\leq \frac{\sqrt{N}\log(2\pi abN)}{2\sqrt{2\pi ab}} + O(L^{-1})
\]
once $L$ is sufficiently large that $\min(|A|,N) = N$. As discussed in Sec.~\ref{appendix:entanglement:density_matrix}, this same conclusion also applies for charges $L-N$, so we conclude that the entanglement grows as an area law for extremal charges.

\subsection{Laplace's method: next-to-leading order\label{appendix:entanglement:laplace2}}
To illustrate numerical agreement at computationally accessible system sizes, we seek to determine $S(\rho_A)$ to next-to-leading order. There is already an $O(1)$ contribution in Eq.~\ref{eq:h0}, but there is an additional $O(1)$ contribution that comes from working to next-to-leading-order in Laplace's approximation. We do so following~\cite{miller2006asymptotics}. In general, Laplace's method gives the following asymptotic expansion for any integral bounds containing $k_*$:
\[
\int \D{k} e^{Lf(k)}\tilde{h}_i(k)e^{r_i(k)} \sim \sqrt{\frac{\pi}{L}}e^{Lf(k_*)}\sum_{j=0}^\infty \frac{\phi^{(2j)}(0)}{2^{2j}j!L^j},\label{eq:full_laplace}
\]
where
\[
\phi_i(s) = \tilde{h}_i(k_*+sv(s))e^{r_i(k_*+sv(s))}(sv(s))'
\]
with $v(s)$ defined such that
\[
\frac{f(k_*+sv(s))}{s^2} &= -1.\label{eq:v_def}
\]
This machinery is set up to transform the exponential in the integrand into a proper Gaussian so that arguments in Eqs.~\ref{eq:laplace1} and \ref{eq:laplace3} are made precise. We can determine $v(s)$ order-by-order by expanding $f(k_*+sv(s))$ about $k_*$. Determining $v(s)$ to zeroth order in $s$ is straightforward:
\[
f(k_*+sv(s)) = \frac{1}{2}f''(k_*)(sv)^2 + O((sv)^3),
\]
so
\[
\frac{1}{2}f''(k_*)v(s)^2 &= -1 + O(sv^3)\\
v(0) &= \sqrt{\frac{-2}{f''(k_*)}}.
\]
Therefore, we find
\[
\phi_i^{(0)}(0) = \tilde{h}_i(k_*)e^{r_i(k_*)}\sqrt{\frac{-2}{f''(k_*)}}
\]
which matches our leading order term from Sec.~\ref{appendix:entanglement:laplace1}. For our next-to-leading order calculation, we need to find $\phi^{(2)}(s)$. First, we note that
\[
\phi_i^{(1)}(s) &= \tilde{h}_i e^{r_i}(sv)''+(\tilde{h}_i e^{r_i})'[(sv)']^2,
\]
where $\tilde{h}_i$ and $r_i$ are evaluated at $k = k_* + sv(s)$.
It follows that:
\[
\phi^{(2)}(s) &= \subalign{& (\tilde{h}_i e^{r_i})(sv)'''\\
&+ 3(\tilde{h}_i e^{r_i})'(sv)'(sv)''\\
&+ (\tilde{h}_i e^{r_i})''[(sv)']^3}\\
\phi^{(2)}(s) &= \subalign{& (\tilde{h}_i e^{r_i})(3v''+sv''')\\
&+ 3(\tilde{h}_i e^{r_i})'(v'+sv)(2v'+sv'')\\
&+ (\tilde{h}_i e^{r_i})''[v+sv']^3}\\
\phi^{(2)}(0) &= \subalign{&3(\tilde{h}_i e^{r_i})|_{s=0}v''(0)\\
&+ 6(\tilde{h}_i e^{r_i})'|_{s=0}v'(0)^2\\
&+(\tilde{h}_i e^{r_i})''|_{s=0}v(0)^3.}
\]
The $\phi^{(2)}(0)$ contribution enters Eq.~\ref{eq:full_laplace} suppressed by a factor of $L^{-1}$, so $O(1)$ contributions will only come from terms in $\phi^{(2)}(0)$ that are $O(L)$.
We have already seen from Eqs.~\ref{eq:h0} and \ref{eq:h1} that $h(k_*)$ and $h'(k_*)$ are $O(\log L)$, and since $g^{(n)}(k_*)$ is $O(1)$ and $r_i(k_*)$ and $r_i'(k_*)$ are $O(L^{-1})$, it follows that $(\tilde{h}_i e^{r_i})$ and $(\tilde{h}_i e^{r_i})'$ are $O(\log L)$ at $k = k_*$. Also, since $v^{(n)}(0)$ can be calculated exclusively in terms of $f^{(n)}(k_*)$ which are independent of system size, $v^{(n)}(0) = O(1)$. Consequently, $(\tilde{h}_i e^{r_i})''v(0)^3$ is the only $O(L)$ contribution to $\phi^{(2)}(0)$. More explicitly,
\[
(\tilde{h}_i e^{r_i})''|_{s=0} = \subalign{&e^{r_i(k_*)}\big(\tilde{h}_i''(k_*)+2\tilde{h}_i'(k_*)r_i'(k_*)\\
&+\tilde{h}_i(k_*)((r_i'(k_*))^2+r_i''(k_*)\big).}
\]

\begin{figure}
    \centering
    \includegraphics[width=\linewidth]{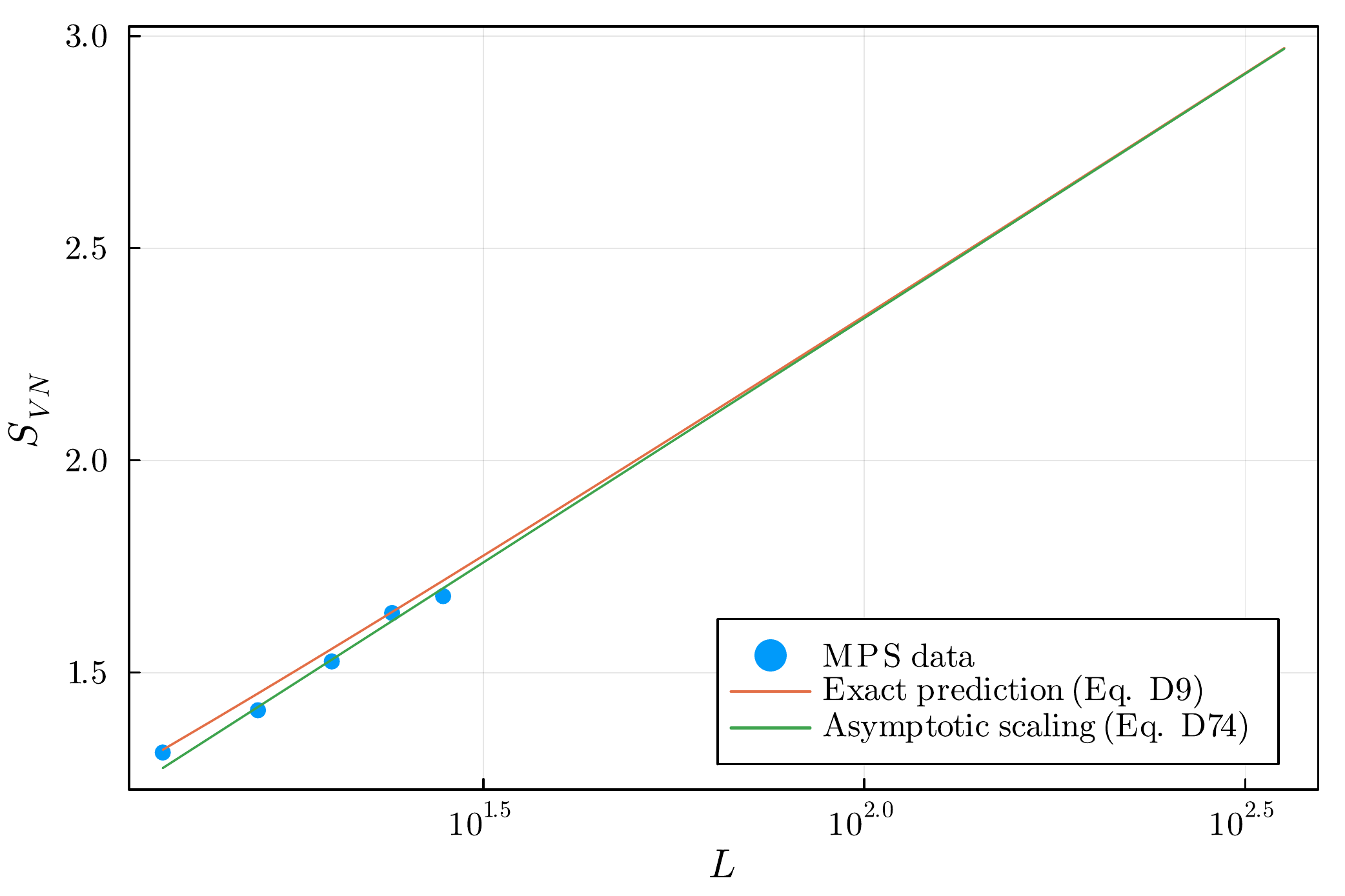}
    \caption{Steady-state half-chain entanglement at half-filling $(a=b=n=\frac{1}{2})$ in the baseline model. Alongside the numerical MPS data are two lines: the exact prediction from Eq.~\ref{eq:entanglement_sum} for the half-chain entanglement of the state in Eq.~\ref{eq:steady_state}, and the asymptotic scaling given in Eq.~\ref{eq:asymptotic_scaling}. We observe that the exact prediction and asymptotic scaling converge for large $L$ and appear consistent with the numerical MPS data.}
    \label{fig:appendix_entanglement_scaling}
\end{figure}

Noting that $r_i''(k_*)$ is also $O(L^{-1})$, we conclude that only $e^{r_i(k_*)}\tilde{h}_i''(k_*)$ can provide $O(L)$ contributions, so that
\[
(\tilde{h}_i e^{r_i})''|_{k=k_*} &= h_i''(k_*) + O(\log L)\\
&= -\frac{L}{(abn(1-n))^{3/2}} + O(\log L).
\]

We find that
\[
v(0)^3 = (2abn(1-n))^{3/2}
\]
so
\[
\phi^{(2)}(0) &= -2^{3/2}L + O(\log L)
\]
and
\[
S(\rho_A) &= \subalign{&\frac{1}{2}(\log L + \log(2\pi abn(1-n)))\\
&+ \frac{1}{\sqrt{2}}\frac{2^{3/2}L}{4L} + O\left(\frac{\log L}{L}\right)}\\
&=  \subalign{&\frac{1}{2}\log L + \frac{1}{2}(\log(2\pi abn(1-n))+1)\\
&+ O\left(\frac{\log L}{L}\right).}\label{eq:asymptotic_scaling}
\]
This asymptotic scaling is compared to the exact result and to numerical data in Fig.~\ref{fig:appendix_entanglement_scaling}, and the leading order contribution agrees with results for the ideal Bose gas~\cite{metlitski2011entanglement}.

\end{document}